\documentclass[%
reprint,
onecolumn,
 amsmath,amssymb,
 aps,
prb,
]{revtex4}

\usepackage{graphics}
\usepackage{dcolumn}
\usepackage{bm}

\usepackage{array}
\usepackage{multirow}
\usepackage{esint}
\usepackage{color}
\usepackage{graphicx}
\usepackage{epstopdf}
\usepackage{verbatim}
\usepackage{float}
\usepackage[caption=false]{subfig}
\usepackage[normalem]{ulem}
\usepackage[colorlinks]{hyperref}

\def\bR{{\mathbf{R}}}
\def\bx{{\mathbf{x}}}
\def\br{{\mathbf{r}}}
\def\bg{{\mathbf{g}}}
\def\R{{\mathbb{R}}}

\newcommand{\angstrom}{\textup{\AA}}

\begin{document}

\title{Symmetry-adapted real-space density functional theory for cylindrical geometries: application to large X (X=C, Si, Ge, Sn) nanotubes }
\author{Swarnava Ghosh}
\affiliation{Division of Engineering and Applied Science, \\ California Institute of Technology. Pasadena, CA 91125 USA}
\author{Amartya S. Banerjee}
\affiliation{Department of Materials Science and Engineering, \\
University of California, Los Angeles, CA 90095 USA}
\author{Phanish Suryanarayana}
\email{phanish.suryanarayana@ce.gatech.edu}
\affiliation{College of Engineering, Georgia Institute of Technology, Atlanta, GA 30332, USA}

\date{\today}

\begin{abstract}
We present a symmetry-adapted real-space formulation of Kohn-Sham density functional theory for cylindrical geometries and apply it to the study of large X (X=C, Si, Ge, Sn) nanotubes. Specifically, starting from the Kohn-Sham equations posed on all of space, we reduce the problem to the fundamental domain by incorporating cyclic and periodic symmetries present in the angular and axial directions of the cylinder, respectively. We develop a high-order finite-difference parallel implementation of this formulation, and verify its accuracy against established planewave and real-space codes. Using this implementation, we study the band structure and bending properties of X nanotubes and Xene sheets, respectively. Specifically, we first show that zigzag and armchair X nanotubes with radii in the range $1$ to $5$ nm are semiconducting, other than the armchair and zigzag type III carbon variants, for which we find a vanishingly small bandgap, indicative of metallic behavior. In particular, we find an inverse linear dependence of the bandgap with respect to the radius for all nanotubes, other than the armchair and zigzag type III carbon variants, for which we find an inverse quadratic dependence. Next, we exploit the connection between cyclic symmetry and uniform bending deformations to calculate the bending moduli of Xene sheets in both zigzag and armchair directions, while considering radii of curvature up to $5$ nm. We find Kirchhoff-Love type bending behavior for all sheets, with graphene and stanene possessing the largest and smallest moduli, respectively. In addition, other than graphene, the sheets demonstrate significant anisotropy, with larger bending moduli along the armchair direction. Finally, we demonstrate that the proposed approach has very good parallel scaling and is highly efficient, enabling ab initio simulations of  unprecedented size for systems with a high degree of cyclic symmetry. In particular, we show that even micron-sized nanotubes can be simulated with modest computational effort. Overall, the current work opens an avenue for the efficient ab-initio study of 1D nanostructures with large radii as well as 1D/2D nanostructures under uniform bending. 
\end{abstract}


\maketitle

\section{Introduction} \label{Section:Introduction}
Over the course of the past few decades, ab-initio calculations based on Kohn-Sham density functional theory (DFT) \cite{Hohenberg,Kohn1965}  have become a mainstay of computational materials research due to their predictive power and ability to provide fundamental insights into materials properties and behavior. Indeed, a relatively large fraction of the computational resources worldwide are now devoted to first principles DFT calculations. The widespread popularity of DFT can be attributed to its generality, relative simplicity, and high accuracy-to-cost ratio when compared to other such ab-initio methods \cite{burke2012dft,becke2014dft}. However, even though the cost of DFT is significantly less than the  more accurate wavefunction-based theories, the efficient solution of the Kohn-Sham equations remains a challenging task, which severely restricts the range of physical systems that can be investigated. In particular, the computational cost and memory storage  scale cubically and quadratically with respect to system size, respectively, and are typically associated with large prefactors. Moreover, the scalability in the context of high performance parallel computing suffers from the global nature of the orthonormality constraint posed on the Kohn-Sham orbitals.

The planewave pseudopotential method \cite{Martin2004} is one of the most popular techniques for performing DFT calculations. In this approach, the Kohn-Sham equations are discretized using the Fourier basis, a complete and systematically improvable set in which convergence can be controlled by a single parameter. The planewave method is therefore not only accurate and simple to use, but is highly efficient on small to moderate computational resources due to the use of optimized Fast Fourier Transforms (FFTs). However, the planewave method suffers from a few limitations, including the following. The Fourier basis enforces periodic boundary conditions, whereby finite systems such as molecules and semi-infinite systems such as nanotubes require the introduction of artificial periodicity, with possibly large  regions of vacuum between periodic replicas. This limitation also requires the introduction of an artificial neutralizing background density when treating charged systems. Furthermore, the global nature of the Fourier basis prevents the development of linear-scaling methods \cite{Goedecker,Bowler2012}. Finally, the reliance on FFTs hampers parallel scaling, thus limiting the length and time scales that can be reached. 

In view of the aforementioned limitations of the planewave method, a number of alternate representations have been developed over the last two decades that are not only systematically improvable but also localized  \cite{arias1999wav,beck2000rsmeth,pask2005femeth,saad2010esmeth,ONETEP,hernandez1997basis,suryanarayana2011mesh,lin2012adaptive}. Among these approaches,  real-space finite-difference methods \cite{beck2000rsmeth,saad2010esmeth}---for which computational locality is maximized by discretizing all quantities of interest on a real-space grid using high-order finite-differences---are some of the most mature and widely used to date. In these methods, convergence is again controlled by a single parameter, i.e., the mesh-size or grid spacing. In addition, any of Dirichlet, periodic, and Bloch-periodic boundary conditions (and combinations thereof) can be accommodated, whereby finite, semi-infinite, and charged systems, as well as bulk 3D systems can all be accurately and efficiently studied. Moreover, the locality of the discretization and decay of the density matrix within this representation \cite{Suryanarayana2017Nearsightedness} allows for the development of linear-scaling methods \cite{MGMolref,suryanarayana2013spectral,Suryanarayana2017SQDFT}. Finally,  large-scale parallel computational resources can be efficiently leveraged by virtue of the method's simplicity, locality, and freedom from communication-intensive transforms such as FFTs. The limitations of real-space methods include the larger number of degrees of freedom/atom and the lack of effective preconditioners, when compared to the planewave method.  

In recent years, there has been a significant increase in the efficiency of real-space methods due to a  number of advances. Since early work in this area \cite{bernolc1991mg,chel1994fdpp,briggs1995mg,seits1995fdcg}, the degrees of freedom/atom  required to obtain accurate ground state properties has been notably reduced by double-grid techniques \cite{OnoHir99}, ultrasoft pseudopotential formulations \cite{hodak2007rsuspp}, projector augmented wave methods \cite{mort2005rspaw}, high-order integration \cite{BobSchChe15}, reformulation of nonlocal pseudopotential components  \cite{hirose2005first,sharma2018calculation}, and reduction of the eigenproblem by discontinuous projection \cite{xu2018discrete}. Moreover, the need for effective preconditioners has been circumvented by substituting traditional iterative eigensolvers with the Chebyshev-polynomial filtered subspace iteration (CheFSI) \cite{zhou2006self}. These and other advances have made large systems containing thousands of atoms amenable to  real-space finite-difference methods \cite{alamany2008fdlarge}. In fact, they are now able to outperform  established planewave codes in the context of both finite \cite{Ghosh2017cluster} and extended \cite{Ghosh2017extended} systems. However, just like planewave methods, real-space methods have been restricted to affine (primarily Cartesian) coordinate systems. Though these are ideally suited for the various crystal systems, curvilinear coordinate systems provide an opportunity for  the better description of systems with curved geometries.  

1D nanostructures possessing a cylindrical-type geometry, e.g., nanotubes, nanowires, and nanorods,  have received a lot of attention in the past three decades due to their unusual and fascinating material properties \cite{Martel1998,javey2003ballistic,popov2004carbon,gong2009nitrogen,park2009silicon,wu2012stable,park2011germanium,li2011controlled,zhao2006porous,patzke2002oxidic,law2004semiconductor}. This is also true for their 2D counterparts \cite{xu2013graphene,bhimanapati2015recent,butler2013progress,naguib201425th,fiori2014electronics,koppens2014photodetectors}, which have risen to prominence after the discovery of graphene \citep{novoselov2004electric}. Though these 2D  structures are originally planar, they take up cylindrical-type geometries when subject to bending deformations, a common and technologically relevant mode of deformation in such materials \cite{schniepp2008bending,wei2012bending}. These cylindrical-type geometries are indeed best described while working in a cylindrical coordinate system. In particular, such a choice enables the implementation to be compatible with the (possibly large) rotational/cyclic symmetry that could be present in the system. Indeed, such symmetry is commonly found in 1D nanostructures, and due to its connection with uniform bending deformations \citep{james2006objective,Banerjee2016cyclic}, also while studying 2D nanostructures subject to bending deformations. However, being restricted to affine coordinate systems, current DFT methods are unable to fully exploit the cyclic symmetry to reduce the computational cost, which is critical for 1D nanostructures such as nanotubes with large radii \cite{shin2004formation,mcgary2006magnetic,macak2008mechanistic} as well as for bending of 2D nanostructures with radii of curvature representative of those realized in experiments. The suitability of the real-space method for cylindrical coordinates and its attractive features outlined above provide the motivation for the current effort. 

In this work, we present a symmetry-adapted real-space formulation of Kohn-Sham DFT for cylindrical geometries and apply it to the study of large X (X=C, Si, Ge, Sn) nanotubes. Specifically, we incorporate the cyclic and periodic symmetries present in the angular and axial directions of the cylinder, respectively, to reduce the Kohn-Sham equations that are originally posed on all of space to the fundamental domain. We develop a parallel implementation of this formulation using high-order finite-differences, and verify its accuracy by benchmarking against established planewave and real-space codes. Using this implementation, we study the band structure properties of X nanotubes and the bending properties of Xene sheets. Specifically, we first show that zigzag and armchair X nanotubes of radii from $1$ to $5$ nm are semiconducting, other than the armchair and zigzag type III carbon variants, for which we find a vanishingly small bandgap, indicative of metallic behavior. In particular, we find an inverse quadratic dependence of the bandgap with respect to radius for armchair and zigzag type III carbon nanotubes, other than which, there is an inverse linear dependence. Next, using the connection between cyclic symmetry and uniform bending deformations, we calculate the bending moduli of Xene sheets in both zigzag and armchair directions, while considering radii of curvature up to $5$ nm. For all sheets, we observe a Kirchhoff-Love type bending behavior, with graphene and stanene possessing the largest and smallest moduli, respectively. In addition, apart from graphene, there is significant anisotropy in the bending moduli, with larger values along the armchair direction. Finally, we demonstrate that the proposed method has very good parallel scaling and is highly efficient, enabling ab initio simulations of  unprecedented size for systems with sufficiently high degree of cyclic symmetry, e.g., we simulate a silicon nanotube of radius $\sim 1 \mu$m within $53$ minutes on $353$ processors.

The remainder of this manuscript is organized as follows. In Section~\ref{Sec:RSDFT}, we summarize the underlying real-space formulation of DFT that will be adopted here. In this framework, we develop a symmetry-adapted formulation of DFT for cylindrical geometries, as described in Section~\ref{Section:Formulation}. We then describe its parallel implementation in Section~\ref{Section:Implementation}. Next, we demonstrate the accuracy and efficiency of the proposed formulation and implementation in Section~\ref{Section:ExamplesResults}, a section in which we also study the electronic properties of X nanotubes and the bending properties of Xene sheets. Finally, we provide concluding remarks in Section~\ref{Sec:Conclusions}.


\section{Real-space formulation of DFT} \label{Sec:RSDFT}
We first provide some mathematical background on Kohn-Sham DFT, while adopting a real-space formalism that has been shown to be both accurate and efficient \cite{Ghosh2017cluster,Ghosh2017extended}. Neglecting spin and utilizing the pseudopotential approximation, the electronic free energy of a system can be written as \citep{Kohn1965, mermin1965thermal}:
\begin{align}
\bar{\mathcal{F}} (\Psi, \bg, \textbf{X}) = \bar{T}_s(\Psi,\bg) & + \bar{E}_{xc}(\rho)  + \bar{K}(\Psi,\bg, \textbf{X}) + \bar{E}_{el}(\rho, \textbf{X}) - T \bar{S}(\bg) \,,
\label{Eqn:free_energy}
\end{align}
where $\Psi$ is the collection of Kohn-Sham orbitals associated with the system, $\bg$ is the corresponding collection of orbital occupation numbers, $ \textbf{X}$ is the collection of atomic positions, and $\rho$ is the electron density. The electron density can itself be expressed in terms of the orbitals and their occupations as:
\begin{align}
\label{Eqn:electron_density}
\rho(\br) = 2 \sum_{n=1}^{\bar{N}_s} g_{n} \lvert{\psi_n(\br)}\rvert^2 \,,
\end{align} 
where $\bar{N}_s$ denotes the total number of orbitals. Above and in what follows, the generic index $n$ is used to label the orbitals and the corresponding orbital occupations (i.e., $\psi_n \in \Psi$ is a typical orbital and its occupation is $g_n \in \bg$), while the generic index $J$ will provide labels for the nuclei (i.e., $\textbf{x}_J \in \textbf{X}$ is the position of the $J^{\text{th}}$ nucleus). In addition, the subscript $J$ will be used to explicitly indicate the dependence of a function (or a collection) on the species of the atom located at $\textbf{x}_J$. 
 
The various terms arising in Eq.~\ref{Eqn:free_energy} can be interpreted as follows. The first term models the kinetic energy of the system of electrons, and it can be written as:
\begin{align}
\label{Eqn:Kinetic_Energy}
\bar{T}_s(\Psi,\bg) = - \sum_{n=1}^{\bar{N}_s} g_{n} \int_{\mathbb{R}^{3}}  \psi^{*}_{n}(\br) \nabla^2   \psi_n(\br)\, \mathrm{d\br} \,.
\end{align}
The second term represents the exchange-correlation energy, for which many models exist, including the Local Density Approximation (LDA)\citep{Kohn1965} and the Generalized Gradient Approximation (GGA) \citep{perdew1996generalized}. In this work, we employ the LDA:
\begin{align}
\label{Eqn:Exc}
E_{xc} (\rho) = \int_{\mathbb{R}^{3}} \varepsilon_{xc} (\rho(\br)) \rho(\br) \, \mathrm{d \br} \,.
\end{align}
The third term accounts for the contribution from the nonlocal part of the pseudopotentials, which takes the following form within the Kleinman-Bylander representation \cite{kleinman1982efficacious}:
\begin{align}
\bar{K}(\Psi,\bg, \textbf{X}) = 2 \sum_{n=1}^{\bar{N}_s} g_{n} \sum_{J} \sum_{p \in \mathcal{A}_{J}} \gamma_{J; p} \left| \int_{\mathbb{R}^3} \chi_{J; p}^{*}(\textbf{x}_J, \textbf{r})\,\psi_n(\textbf{r}) \, \mathrm{d\textbf{r}} \, \right|^2\,.
\label{Eqn:Nonlocal}
\end{align}
Here, $\mathcal{A}_{J}$ denotes the collection of projectors associated with the atom at $\textbf{x}_J$,  and $\chi_{J; p}$ are the nonlocal projection functions, with $\gamma_{J; p}$ representing the corresponding normalization constants. The fourth term represents the total electrostatic interaction energy, for which we use a local formulation of the electrostatics \citep{Pask2005,Suryanarayana2014524,ghosh2014higher} suitable for real-space DFT: 
\begin{align}  
\bar{E}_{el} (\rho, \textbf{X}) =\max_{\phi}  \bigg \{ - \frac{1}{8 \pi} \int_{\mathbb{R}^3} |\nabla \phi(\br)|^2 \, \mathrm{d\br} + \int_{\mathbb{R}^3}\big(\rho(\br)+ b(\br, \textbf{X}) \big) \phi(\br) \, \mathrm{d\br} \bigg \} + \bar{E}_{sc}( \textbf{X})\,. \label{Eqn:Electrostatics} 
\end{align}
Above, $\phi$ is the electrostatic potential, $b$ is the total pseudocharge density of the nuclei, and $\bar{E}_{sc}$ is the sum of the self energy and the repulsive energy corrections associated with the pseudocharges. Finally, the last term accounts for the electronic entropy arising from fractionally occupied electronic states at a given electronic temperature $T$:
\begin{align}
\label{Eqn:Electronic_Entropy}
\bar{S}(\bg) = -2 k_{B} \sum_{n=1}^{\bar{N}_s} \big( g_{n} \log g_{n} + (1 - g_{n}) \log (1-g_{n}) \big) \,,
\end{align}
where $ k_{B}$ is the Boltzmann constant. 

The computation of the electronic ground state corresponding to the given set of (fixed) atomic positions $\textbf{X}$ is given by the variational problem:
\begin{align} 
 \bar{\mathcal{F}}_0(\textbf{X}) & = \min_{ \Psi ,\bg} \bar{\mathcal{F}}(\Psi,\bg,\textbf{X}) \label{Eqn:GroundStateElectronic} \\    
\mathrm{s.t.} \quad \int_{\R^3} \psi_i(\bx) \psi_j(\bx) \, \mathrm{d\bx} & = \delta_{i,j} \,, \,\,\, i, j = 1, 2, \ldots, \bar{N}_s \,; \,\, \mathrm{and} \,\, 2 \sum_{n=1}^{\bar{N}_s} g_n = \bar{N}_e  \,, \nonumber
\end{align}
where $\delta_{i,j}$ is the Kronecker delta function and $\bar{N}_e $ is the total number of electrons. Note that the constraint of orthonormality  on the orbitals is a consequence of the Pauli exclusion principle and the constraint on the occupations arises from the constancy of the number of electrons due to the Aufbau principle \cite{NumAnalysis2003}. In practice, the electronic ground state is often determined by seeking stationary states corresponding to solutions of the Euler-Lagrange equations:
\begin{align}
\left(\bar{\mathcal{H}}(\Psi, \bg, \textbf{X}) \equiv -\frac{1}{2} \nabla^2 + V_{xc} + \phi +\bar{\mathcal{V}}_{nl} \right) \psi_n &= \lambda_n \psi_n \,, \quad n=1,2,\ldots, N_s \,,
\label{Eqn:eigenvalue_problem}
\end{align}
where $ V_{xc} = \displaystyle \frac{\delta \bar{E}_{xc}}{\delta \rho}$ is the exchange-correlation potential, the electrostatic potential $\phi$ is the solution of the Poisson equation:
\begin{align}
\label{Eqn:poisson_problem}
-\frac{1}{4 \pi} \nabla^2 \phi(\br,\textbf{X}) = \rho(\br)+b(\br, \textbf{X}) \,,
\end{align}
the occupations $g_n$ are given by the Fermi-Dirac function:
\begin{align}
\label{Eqn:occupation_Fermi_Dirac}
g_{n} = \bigg( 1 + \exp{\bigg( \frac{\lambda_n - \lambda_F}{k_B T} \bigg)} \bigg)^{-1}\,, 
\end{align}
and $\bar{\mathcal{V}}_{nl}$---projection operator associated with the non-local part of the pseudopotential---acts on any given function $f(\br)$ as:
\begin{align}
\label{Eqn:VNL_Operator}
[\bar{\mathcal{V}}_{nl} f ](\br) =\sum_{J} \sum_{p \in \mathcal{A}_{J}}\gamma_{J;p}\;\chi_{J;p}(\bx_J, \br)\int_{\mathbb{R}^3}\chi_{\bx;p}^{*}(\bx, \textbf{y}) f(\textbf{y})\, \mathrm{d\textbf{y}}\,.
\end{align}
Note that while solving Eqs.~\ref{Eqn:eigenvalue_problem} and \ref{Eqn:poisson_problem}, the boundary conditions prescribed on the orbitals $\psi_n$ and the electrostatic potential $\phi$ are that they decay to zero at infinity \citep{wavefunc_decay1,wavefunc_decay2,Phanish2010}. Here and henceforth, we assume that the systems are charge neutral. 

Once the electronic ground-state has been determined, the free energy can be calculated using either Eq.~\ref{Eqn:free_energy} or the Harris-Foulkes \cite{harris1985simplified,foulkes1989tight} type functional : 
\begin{align} 
\bar{\mathcal{F}}_0(\textbf{X}) =  2 \sum_{n=1}^{\bar{N}_s}  g_n \lambda_n + \bar{E}_{xc}(\rho)  - \int_{\R^3} V_{xc}(\rho(\br))\rho(\br)\, \mathrm{d\br}+  \frac{1}{2} \int_{\R^3}\big(b(\br,\mathbf{X})-\rho(\br)\big) \phi(\br,\mathbf{X}) \, \mathrm{d\br} + \bar{E}_{sc}(\mathbf{X}) -T\bar{S}(\bg)  \label{Eqn:Harris_Foulkes}
\,,
\end{align}
while the Hellmann-Feynman atomic forces required for geometry optimization and molecular dynamics can be written as:
\begin{eqnarray}
\mathbf{f}_J & = & -\frac{\partial \bar{\mathcal{F}}_0(\textbf{X})}{\partial \textbf{x}_J} \nonumber \\ 
  & = &  \int_{\R^3} \nabla b_{J}(\br, \textbf{x}_J) \phi(\br,\textbf{X}) \, \mathrm{d \br} + \bar{\mathbf{f}}_{sc,J}(\textbf{X})  \nonumber \\
 & - & 4 \sum_{n=1}^{\bar{N}_s} g_{n} \sum_{p \in \mathcal{A}_{J}} \gamma_{J;p}  \text{Re} \left[ \left( \int_{\R^3} \psi_{n}^*(\br) \chi_{J;p}(\bx_J, \br) \, \mathrm{d\br} \right) \left( \int_{\R^3} \nabla \psi_{n}(\br) \chi_{J;p}^*(\bx_J, \br) \, \mathrm{d\br} \right) \right] \,, \label{Eqn:Force:Nuclei}
\end{eqnarray}
where $b_J$ represents the pseudocharge of the $J^{th}$ nucleus and $\text{Re}[.]$ denotes the real part of the bracketed expression. The first term is the local component of the force, the second term $\bar{\mathbf{f}}_{sc,J}(\textbf{X})= \frac{\partial \bar{E}_{sc}(\mathbf{X})}{\partial \bx_J}$ corrects for overlapping pseudocharge densities, and the final term is the nonlocal component of the force. 

\section{Symmetry-adapted real-space DFT on a cylinder} \label{Section:Formulation}
We now present a real-space formulation  of DFT for cylindrical geometries that is able to explicitly incorporate the natural symmetries commonly arising in such systems. In order to achieve this, the Kohn-Sham equations posed on all of space in the previous section will be appropriately modified and augmented with suitable boundary conditions, as detailed below. 

\subsection{System specification: domain, atomic configuration and symmetries}
\begin{figure}[htb]\centering
\includegraphics[width=0.8\textwidth]{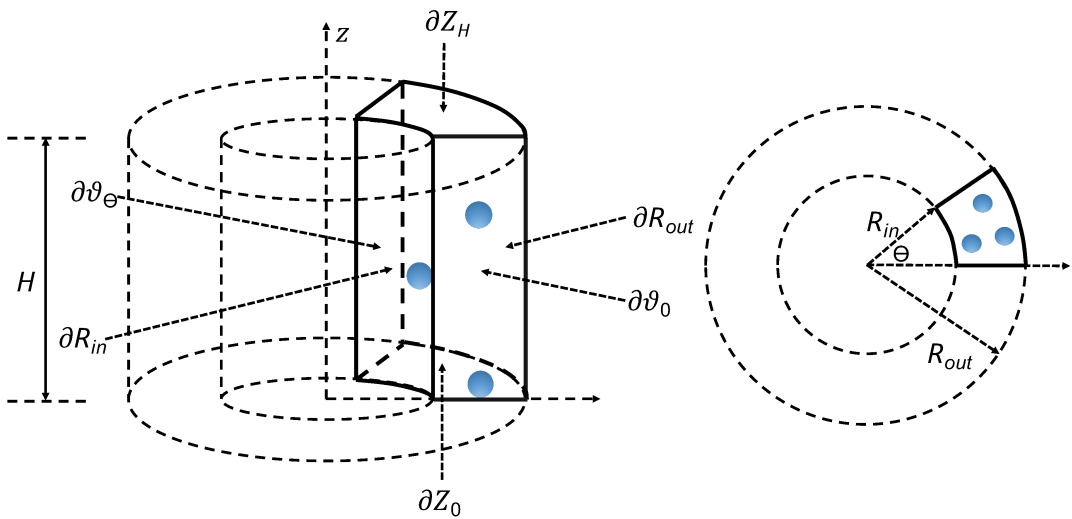}
\caption{Schematic of the annular cylindrical region $\Omega$ with boundary $\partial \Omega$. The left figure represents the side view and the right figure represents the top view. }
\label{Fig:Fundamental_Domain}
\end{figure}

Let $(\textbf{e}_x, \textbf{e}_y, \textbf{e}_z)$ denote the canonical Cartesian coordinate axes and let $(r,\theta,z)$ represent the cylindrical coordinates associated with a point having Cartesian coordinates $(x,y,z)$. Consider an annular cylindrical region:
\begin{equation} \label{Eqn:Omega}
\Omega := \left \{ (r,\theta,z)\in \R^3 \big \vert R_{in}\leq r \leq R_{out}, 0 \leq \theta \leq \Theta \leq 2\pi, 0 \leq z \leq H  \right\} \,,
\end{equation}
with axis along $\textbf{e}_z$. The boundary of $\Omega$ can be decomposed as 
\begin{equation} \label{Eqn:OmegaBoundary}
\partial \Omega= \partial R_{in} {\,\textstyle\bigcup\,} \partial R_{out}  {\,\textstyle\bigcup\,}  \partial \vartheta_{0}  {\,\textstyle\bigcup\,}  \partial \vartheta_{\Theta}  {\,\textstyle\bigcup\,}  \partial\mathcal{Z}_0 {\,\textstyle\bigcup\,}  \partial{\mathcal{Z}}_H \,,
\end{equation}
where $\partial R_{in}$ and $\partial R_{out}$ denote the surfaces $r = R_{in}$ and  $r = R_{out}$, respectively; $\partial \vartheta_{0}$ and $\partial \vartheta_{\Theta}$ denote the surfaces $\theta = 0$ and $\theta = \Theta$, respectively; and $\partial\mathcal{Z}_0$ and $\partial{\mathcal{Z}}_H$ denote the surfaces $z = 0$ and $z= H$, respectively. Let this region $\Omega$ contain $N$ atoms positioned at $\bR_1, \bR_2, \ldots, \bR_N$, the collection of which will henceforth be denoted by $\bR$. A schematic illustrating this system is as shown in Fig.~\ref{Fig:Fundamental_Domain}.

The system under consideration is allowed to have transational symmetry along the axis $\textbf{e}_z$ and/or rotational symmetry in the angular direction, i.e., cyclic symmetry about axis $\textbf{e}_z$. In the former case, the height of the domain $H$ must be commensurate with the translational periodicity of the system along $\textbf{e}_z$. In the latter, the maximum polar angle $\Theta$ associated with the domain must be of the form $\Theta = 2\pi / \mathfrak{N}$, with $\mathfrak{N}$ being a natural number. In such cases, the region $\Omega$ is a \textit{fundamental domain} or the \emph{unit cell} for the symmetries involved, which plays a role similar to that of the unit cell employed in standard periodic calculations. In the current context, whenever periodic symmetries are present, the system under study is quasi-1D in nature. The symmetry group in this case is generated by translations along $\textbf{e}_z$, and is of the form:
\begin{equation}
\mathfrak{T}=\{\textbf{t}_{\mu}: \mu \in \mathbb{Z}\} \,, \,\, \text{with} \,\,  \textbf{t}_{\mu}=\mu H \textbf{e}_z\,.
\end{equation} 
When cyclic symmetries are present, the symmetry group is generated by rotations with the common axis $\textbf{e}_z$, and it can be written as the set of matrices 
\begin{align}
\label{Eqn:rot_mat}
\mathfrak{C}=\{\mathfrak{R}^{\zeta}: \zeta = 0,1,2,\ldots,\mathfrak{N}-1\} \,, \,\, \text{with} \,\, \mathfrak{R}^{\zeta} =
\begin{pmatrix}
\cos \frac{2\pi \zeta}{\mathfrak{N}}   & -\sin \frac{2\pi \zeta}{\mathfrak{N}} & 0 \vspace{0.05in}\\
\sin \frac{2\pi \zeta}{\mathfrak{N}} & \cos \frac{2\pi \zeta}{\mathfrak{N}} & 0 \vspace{0.05in} \\
0 & 0 & 1
\end{pmatrix}\,.
\end{align}

The symmetry groups associated with the (tubular) physical systems studied in this work are a combination of cyclic and translational symetries, i.e., the group in such cases is expressible as the direct product of the groups $\mathfrak{T}$ and $\mathfrak{C}$. Thus, the group  is identifiable as the set of isometries (i.e., rigid body motions):
\begin{align}
\label{Eqn:isometry}
\mathcal{G} = \{{\Upsilon}_{\zeta, \mu} = (\mathfrak{R}^{\zeta}| \textbf{t}_{\mu} ): \mathfrak{R}^{\zeta} \in \mathfrak{C},  \textbf{t}_{\mu} \in \mathfrak{T}\}\,,
\end{align}
and it can be indexed by pairs of numbers $(\zeta,\mu)$ with $\zeta = 0,1,2,\ldots,\mathfrak{N}-1$ and $\mu \in \mathbb{Z}$. Specifically, the group element associated with the pair $(\zeta,\mu)$ is the isometry ${\Upsilon}_{\zeta, \mu}=(\mathfrak{R}^{\zeta}| \textbf{t}_{\mu} )$, whose action on a point in space (denoted henceforth as $\circ$) rotates it by $\mathfrak{R}^{\zeta}$ about $\textbf{e}_z$, while also simultaneously translating it by $\mu H$ along $\textbf{e}_z$. The global tubular structure $\textbf{X}$ that is effectively being simulated can be generated as the image of the points $\bR_1, \bR_2, \ldots, \bR_N$ under the action of the isometries in  the group $\mathcal{G}$, i.e., 
\begin{align}
\textbf{X} = \{ {\Upsilon}_{\zeta, \mu}{\circ}\bR_{J} = \mathfrak{R}^{\zeta} \bR_J  + \textbf{t}_{\mu} \}\,, \text{with}\;J = 1,2,\ldots, N\,, \,\, \zeta = 0,1,2,\ldots,\mathfrak{N}-1\,, \,\, \mu \in \mathbb{Z}\,.
\label{Eqn:full_structure}
\end{align}
Correspondingly, the global simulation domain $\mathcal{C}$, that encases all of the points in $\textbf{X}$, can be generated as the image of the fundamental domain $\Omega$ under the group $\mathcal{G}$. We will use the notation $\mathcal{G} \circ \bR_J$ to denote the orbit of the point $\bR_j$ under the group, i.e., 
\begin{align}
\label{Eqn:orbit_def}
\mathcal{G} \circ \bR_J = \big\{\Upsilon_{\zeta,\mu}\circ \bR_J: \Upsilon_{\zeta,\mu} \in \mathcal{G}\big\}\,.
\end{align}
With this notation, the collection of points $\textbf{X}$ is expressible as:
\begin{align}
\textbf{X} = \bigcup_{J=1}^N \mathcal{G} \circ{\bR_J}\,.
\end{align} 
Indeed, in the context of DFT, $\textbf{X}$ is nothing but the collection of atomic positions in $\R^3$. 
 

\subsection{Formulation: Symmetry-adapted real-space DFT }
A basic consequence of the presence of physical symmetries in a system---specifically, the atomic positions of the structure being describable as the orbit of a discrete group of isometries---is that, under some generally applicable hypotheses \citep{Banerjee_PhD_Thesis, Banerjee2016cyclic}, the electron density for such a system is invariant under the symmetry group and further, the Kohn-Sham Hamiltonian for the system commutes with the symmetry operations of the group. Results from group representation theory dictate therefore, that the eigenstates of this operator can be characterized through the irreducible representations of the symmetry group, and that the eigenstates transform as the irreducible representations under the action of the group\citep{Banerjee_PhD_Thesis, mcweeny2002symmetry, hamermesh2012group}. The relevant symmetry group $\mathcal{G}$ in the present context is Abelian, therefore, its complex irreducible representations are one dimensional \citep{ Folland_Harmonic, Barut_Reps}. These one-dimensional irreducible representations, or the so called \textit{complex characters} of $\mathcal{G}$ are complex valued functions of the group: identifying the group element $\Upsilon_{\zeta,\mu}  \in \mathcal{G}$ in terms of the pair $(\zeta,\mu) \in \{0,1,2,\ldots,\mathfrak{N} - 1\} \times \mathbb{Z}$, the set of characters of $\mathcal{G}$ can be written as:
\begin{align}
\label{Eqn:dual_group}
\widehat{\mathcal{G}} =\bigg\{e^{2\pi i \big(\frac{\zeta}{\mathfrak{N}}{\nu} + \frac{\mu H}{2\pi} {\eta} \big)} :\,&{\nu} \in  \{0,1,2,\ldots,\mathfrak{N} - 1\}; {\eta} \in \left[-\frac{\pi}{H}, \frac{\pi}{H} \right] \bigg\}\,.
\end{align}
The variables ${\nu}$ and ${\eta}$ serve to label the complex characters of $\mathcal{G}$, and consequently, they also label the associated eigenstates of the Kohn-Sham Hamiltonian. In what follows, we explicitly indicate this labeling for eigenvalues, eigenvectors, and occupations as $\lambda_n(\nu,\eta), \psi_n(\br, \nu,\eta)$, and $g_n( \nu,\eta)$ respectively. 

A number of consequences are associated with, or result from the above observations, as we now discuss. First, due to the orthogonality relations obeyed by the characters\citep{Folland_Harmonic, Barut_Reps}, the collections of eigenstates associated with distinct characters are mutually orthogonal. Hence, by the use of a symmetry adapted basis \citep{mcweeny2002symmetry}, the Hamiltonian $\bar{\mathcal{H}}$ can be block-diagonalized\citep{Banerjee_PhD_Thesis} and the eigenvalue problems associated with distinct characters (i.e., distinct values of $(\eta, \nu)$) can be solved independently of one another. Second, the boundary conditions on the orbitals, and the electrostatic potential, that have to be applied on certain surfaces of the computational domain $\Omega$ can be readily deduced based on transformation properties of the characters. Third, any quantity that involves contributions from all eigenstates that appear in the problem, has to include contributions from each of the elements of $\widehat{\mathcal{G}}$ -- this can be achieved by integrating relevant eigenstate-dependent quantities against a suitable integration measure over $\widehat{\mathcal{G}}$. As an example, consider the electron density: if each of the diagonal blocks of the symmetry-adapted Hamiltonian contributes $N_s$ electronic states, Eq.~\ref{Eqn:electron_density} can be rewritten as:
\begin{align}
\label{Eqn:Sym_electron_density_1}
\rho(\br) = 2 \sum_{n=1}^{N_s} \bigg(\frac{1}{\mathfrak{N}} \sum_{\nu=0}^{\mathfrak{N}-1} \fint  g_{n}(\nu,\eta)\,| \psi_{n}(\br,\nu,\eta) |^2 \, \mathrm{d\eta}\,\bigg).
\end{align}
Here, the sum $\displaystyle \frac{1}{\mathfrak{N}} \sum_{\nu=0}^{\mathfrak{N}-1} $ is associated with integrating against $\nu$, and $\displaystyle \fint$, which signifies the average over the interval $\displaystyle \left[-\frac{\pi}{H},\frac{\pi}{H} \right]$, accumulates contributions in $\eta$. We now exploit these consequences to reduce the Kohn-Sham problem described in Section~\ref{Sec:RSDFT} to the \emph{fundamental domain} $\Omega$. 

\subsubsection{Boundary conditions}

\paragraph{Orbitals} The eigenfunctions of the Hamiltonian transform in accordance with the irreducible representations, whereby the action of an arbitrary group element $\Upsilon_{\zeta,\mu}  \in \mathcal{G}$ on an orbital associated with the character $(\nu, \eta) $ can be written as:
\begin{align}
\label{Eqn:Bloch_theorem_1}
\psi_n(\Upsilon_{\zeta,\mu}^{-1}\circ\br,\nu,\eta) = e^{2\pi i \big(\frac{\zeta}{\mathfrak{N}}{\nu} + \frac{\mu H}{2\pi} {\eta} \big)}\psi_n(\br,\nu,\eta)\,,
\end{align}
or equivalently
\begin{align}
\label{Eqn:Bloch_theorem_2}
\psi_n(\Upsilon_{\zeta,\mu}\circ\br,\nu,\eta) = e^{-2\pi i \big(\frac{\zeta}{\mathfrak{N}}{\nu} + \frac{\mu H}{2\pi} {\eta} \big)}\psi_n(\br,\nu,\eta)\,.
\end{align}
These can be identified as versions of the Bloch-theorem\citep{bloch1929quantenmechanik, Odeh_Keller} associated with the symmetry group $\mathcal{G}$. Therefore, we arrive at the following boundary conditions for the orbitals on the surfaces $\partial \vartheta_{0}  {\,\textstyle\bigcup\,}  \partial \vartheta_{\Theta}$ and $\partial\mathcal{Z}_0 {\,\textstyle\bigcup\,}  \partial{\mathcal{Z}}_H$, respectively:
\begin{align}
\label{Eqn:psi_theta_BC}
& \psi_n(r, \theta = \Theta, z, \nu, \eta) = e^{-\frac{2\pi i \nu}{\mathfrak{N}}}\psi_n(r, \theta = 0, z, \nu, \eta)\,,\\
& \psi_n(r, \theta, z = H, \nu, \eta) = e^{-i \eta H} \psi_n(r, \theta, z = 0, \nu, \eta)\,\label{Eqn:psi_Z_BC} \,.
\end{align}
For the surfaces $\partial R_{in}$ and $\partial R_{out}$, we assume that the atoms within $\Omega$ are sufficiently far from these surfaces, allowing the decay of the electron density along the radial direction to come into effect, i.e., 
\begin{align}
\psi_n(r = R_{in}, \theta , z, \nu, \eta) = \psi_n(r = R_{out}, \theta, z, \nu, \eta) = 0\,. \label{Eqn:psi_R_BC}
\end{align}

\paragraph{Electrostatic potential} The electron density $\rho$ is group invariant, i.e., it is transforms under the group as functions associated with the characters $(\nu,\eta) = (0,0)$. It can be easily shown that the total pseudocharge density $b$ inherits this symmetry as well \citep{Banerjee2016cyclic}, since the atomic positions are expressible as the orbit of the group and the individual pseudocharges are spherically symmetric. Therefore, it follows from Eq.~\ref{Eqn:poisson_problem} that $\phi$ is group invariant, which implies that on $\partial \vartheta_{0}  {\,\textstyle\bigcup\,}  \partial \vartheta_{\Theta}$ and $\partial\mathcal{Z}_0 {\,\textstyle\bigcup\,}  \partial{\mathcal{Z}}_H$, respectively, we have:
\begin{align}
\label{Eqn:phi_theta_BC}
\phi(r, \theta = \Theta, z, \bR, \mathcal{G}) = \phi(r, \theta = 0, z, \bR, \mathcal{G})\,, \\
\phi(r, \theta, z = H, \bR, \mathcal{G}) = \phi(r, \theta, z = 0, \bR, \mathcal{G})\,.
\label{Eqn:phi_Z_BC}
\end{align}
For the surfaces $\partial R_{in}$ and $\partial R_{out}$, the boundary conditions can be determined by using the integral form \citep{Evans_PDE} of the solution to Eq.~\ref{Eqn:poisson_problem}:
\begin{align}
\phi(\br, \textbf{R}, \mathcal{G}) = \int_{\mathbb{R}^3} \frac{\rho(\textbf{y}) + b(\textbf{y}, \textbf{S})}{\lvert \br - \textbf{y}\rvert }\,\mathrm{d\textbf{y}}= \sum_{\Upsilon_{\zeta, \mu} \in \mathcal{G}} \int_{\Omega} \frac{\rho(\textbf{y}) + b(\textbf{y},\textbf{S})}{\lvert \br - \Upsilon_{\zeta, \mu}^{-1}\circ \textbf{y}\rvert }\, \mathrm{d\textbf{y}}\,, \label{Eqn:phi_R_BC}
\end{align}
which can then be evaluated using Ewald summation\citep{miller2010ewald, LANGRIDGE200178} or multipole expansion\citep{han2008real, Ghosh2017cluster} techniques.


\subsubsection{Energy, Kohn-Sham equations, and atomic forces}
In the discussion that follows, we denote the collection of character dependent electronic states by $\Psi(\widehat{\mathcal{G}})$ and the corresponding collection of electronic occupations by $\bg(\widehat{\mathcal{G}})$. If, as before, each diagonal block of the Kohn-Sham Hamiltonian contributes $N_s$ states, we have:
\begin{align}
\nonumber
\Psi(\widehat{\mathcal{G}}) = \bigg\{\psi_n(\br, \eta, \nu): n &=1,\ldots, N_s,
\nu \in  \{0,1,2,\ldots,\mathfrak{N} - 1\}, {\eta} \in \left[-\frac{\pi}{H}, \frac{\pi}{H} \right] \bigg\}\,.
\end{align}

\paragraph{Energy functional} The presence of symmetries make the global system extended in nature, i.e., the atoms within the simulation domain $\Omega$ represent only a portion of the global infinite structure. Therefore, the energies have to be interpreted in a \textit{per fundamental domain} sense. Moreover, although the computation is confined to the fundamental domain, the various terms in Eq.~\ref{Eqn:free_energy} have to be suitably modified to account for (i) the effect of atoms which belong to the global structure but lie outside the fundamental domain, and (ii) the (possibly infinite) multiplicities of electronic states arising due to the effects of symmetry and the extended nature of the system. Keeping these in mind, the symmetry-adapted electronic free energy per fundamental domain can be written as 
\begin{align}
\mathcal{F}\big(\Psi(\widehat{\mathcal{G}}) , \bg(\widehat{\mathcal{G}}), \textbf{R}, \mathcal{G}\big) = T_s&(\Psi(\widehat{\mathcal{G}}) ,\bg(\widehat{\mathcal{G}})) + E_{xc}(\rho)
+ K(\Psi(\widehat{\mathcal{G}}),\bg(\widehat{\mathcal{G}}), \textbf{R}, \mathcal{G}) 
+ E_{el}(\rho, \textbf{R}, \mathcal{G})) - T S(\bg(\widehat{\mathcal{G}})) \,,
\label{Eqn:free_energy_per_FD}
\end{align}
whose terms are now described in detail.

The first term in the energy functional (Eq.~\ref{Eqn:free_energy_per_FD}) is the electronic kinetic energy per fundamental domain, and like the electron density, it includes contribution from $N_s$ electronic states from each diagonal block of the Hamiltonian. As a result, Eq.~\ref{Eqn:Kinetic_Energy} is modified to:
\begin{align}
\label{Eqn:kinetic_energy_per_FD}
T_s(\Psi(\widehat{\mathcal{G}}) ,\bg(\widehat{\mathcal{G}}))  = - \sum_{n=1}^{N_s}\bigg(\frac{1}{\mathfrak{N}} \sum_{\nu=0}^{\mathfrak{N}-1} \fint \int_{\Omega} {g_{n}(\nu,\eta)}\,\psi_{n}^*(\br,\nu,\eta) \nabla^2 \psi_n(\br,\nu,\eta) \, \mathrm{d \br} \, \mathrm{d\eta}\bigg)\,.
\end{align}

The second term in the energy functional (Eq.~\ref{Eqn:free_energy_per_FD}) is the exchange-correlation energy per fundamental domain, and since the electron density obeys the symmetry of the structure, Eq.~\ref{Eqn:Exc} reduces to:
\begin{align}
\label{Eqn:Exc_per_FD}
E_{xc}(\rho) = \int_{\Omega} \varepsilon_{xc} (\rho(\br)) \rho(\br) \, \mathrm{d \br} \,.
\end{align}
Note that even though we are focusing on LDA in this work, an analogous expression involving the gradient of the electron density is applicable for semilocal exchange-correlation functionals such as the GGA. 

The third term in the energy functional (Eq.~\ref{Eqn:free_energy_per_FD}) is the nonlocal pseudopotential energy per fundamental domain. To obtain this term from Eq.~\ref{Eqn:Nonlocal}, we include contribution from $N_s$ electronic states from each diagonal block of the Hamiltonian to arrive at:
\begin{align}
K(\Psi(\widehat{\mathcal{G}}),\bg(\widehat{\mathcal{G}}), \textbf{R}, \mathcal{G}) = 2 \sum_{n=1}^{N_s}\bigg(\frac{1}{\mathfrak{N}}\sum_{\nu=0}^{\mathfrak{N}-1}\fint {g_{n}(\nu,\eta)}\sum_{J=1}^{N}\sum_{p\in \mathcal{A}_J} \! \gamma_{J;p}\bigg\lvert\int_{\mathcal{C}} \chi^{*}_{J;p}(\bR_J,\br) \psi_{n}(\br,\nu,\eta)  d\br \bigg\rvert^{2} \mathrm{d\eta} \bigg)\,.
\label{Eqn:NLE_per_FD_1}
\end{align} 
Here, we have accumulated the  contribution of the projectors centered on the atoms within the fundamental domain, as applied to all the electronic states in the system. Note that since the atom centered projectors can have support extending beyond the fundamental domain, their overlaps with the orbitals need to be carried out over the global simulation domain $\mathcal{C}$. Eq.~\ref{Eqn:NLE_per_FD_1} can now be rewritten as follows
\begin{align}
K(\Psi(\widehat{\mathcal{G}}),\bg(\widehat{\mathcal{G}}), \textbf{R}, \mathcal{G}) &=2 \sum_{n=1}^{N_s}\bigg(\frac{1}{\mathfrak{N}}\sum_{\nu=0}^{\mathfrak{N}-1}\fint {g_{n}(\nu,\eta)}\sum_{J=1}^{N}\sum_{p\in \mathcal{A}_J} \! \gamma_{J;p}\bigg\lvert \sum_{\Upsilon_{\zeta,\mu} \in \mathcal{G}} \int_{\Upsilon_{\zeta,\mu}\circ\Omega}\!\chi^{*}_{J;p}(\bR_J,\br) \psi_{n}(\br,\nu,\eta)  \mathrm{d\br} \bigg\rvert^{2} \mathrm{d\eta} \bigg)
\nonumber \\
&=2\sum_{n=1}^{N_s}\bigg(\frac{1}{\mathfrak{N}}\sum_{\nu=0}^{\mathfrak{N}-1}\fint {g_{n}(\nu, \eta)} \sum_{J=1}^{N}\sum_{p\in \mathcal{A}_J} \! \gamma_{J;p}\,\bigg\lvert \sum_{\Upsilon_{\zeta,\mu} \in \mathcal{G}} \int_{\Omega} \chi^{*}_{J;p}(\bR_J,\Upsilon_{\zeta,\mu}\circ\textbf{y})  e^{-2\pi i \big(\frac{\zeta}{\mathfrak{N}}{\nu} + \frac{\mu H}{2\pi} {\eta} \big)}\psi_{n}(\textbf{y},\nu,\eta) \, \mathrm{d\textbf{y}} \bigg\rvert^{2} \mathrm{d\eta} \bigg) \nonumber \\
&= 2 \sum_{n=1}^{N_s}\bigg(\frac{1}{\mathfrak{N}}\sum_{\nu=0}^{\mathfrak{N}-1}\fint\,g_{n}(\nu,\eta) \sum_{J=1}^{N} \sum_{p \in \mathcal{A}_J}\gamma_{J;p} \bigg \lvert \int_{\Omega} \widetilde{\chi}^{*}_{J;p}(\bR_J, \br, \nu, \eta)\,\psi_{n}(\br, \nu,\eta)\, \mathrm{d\br} \bigg\rvert^2 \mathrm{d\eta} \bigg)\,,
\label{Eqn:NLE_per_FD}
\end{align}
where the function:
\begin{align}
\widetilde{\chi}_{J;p}(\bR_J, \br, \nu, \eta) = 
\sum_{\substack{\bR_{J'} =  \Upsilon_{\zeta, \mu} \circ \bR_{J}\\ \Upsilon_{\zeta, \mu} \in \mathcal{G}}}\!{\chi}_{J;p}(\bR_{J}, \mathfrak{R}^{\zeta} \br) e^{i( \nu(\theta_J-\theta_{J'})+ \eta (z_J - z_{J'}))} \,.
\label{Eqn:Projector_Structure_Factor} 
\end{align}
The first equality in Eq.~\ref{Eqn:NLE_per_FD} has been obtained by expressing the the global simulation domain $\mathcal{C}$ as the image of the fundamental domain $\Omega$ under the action of the group $\mathcal{G}$. The second equality is obtained by employing the change of variables $\br = \Upsilon_{\zeta,\mu} \circ \textbf{y}$ and then using Eq.~\ref{Eqn:Bloch_theorem_2}. To obtain the third equality, it should be noted that the atomic position $\bR_J$ can be expressed as $\Upsilon_{\zeta,\mu}\circ\bR_{J'}$, with $\bR_{J'}$ denoting an image of $\bR_J$ that (in general) lies away from the fundamental domain. Correspondingly, the quantities $\displaystyle \frac{-2\pi\zeta}{\mathfrak{N}}$ and $-\mu H$ that appear in the exponential can be expressed in terms of the differences in polar coordinates $(r_J, \theta_J, z_J)$ and $(r_{J'}, \theta_{J'}, z_{J'})$ of the atomic positions $\bR_J$ and $\bR_{J'}$. In particular, we have used \citep{Helical_DFT_paper} the fact that the projectors $\chi_{J,p}$ have a mathematical form that is similar to atomic orbitals, i.e, a spherical harmonic multiplied by a compactly supported radial function. 

The fourth term in the energy functional (Eq.~\ref{Eqn:free_energy_per_FD}) is  the electrostatic interaction energy per fundamental domain. Since all quantities involved obey the symmetry of the structure, it is sufficient to work with their restriction to the fundamental domain. Therefore, Eq.~\ref{Eqn:Electrostatics} reduces to:
\begin{align}
E_{el}(\rho, \textbf{R}, \mathcal{G}) = \max_{{\phi}}  \bigg \{ - \frac{1}{8 \pi} \int_{\Omega} \big\lvert\nabla{\phi}(\br, \mathcal{G}) \big \rvert^2 \, \mathrm{d\br}+ \int_{\Omega}\big(\rho(\br)+ b(\br, \textbf{R}, \mathcal{G}) \big)\,{\phi}(\br, \mathcal{G}) \, \mathrm{d\br} \bigg \} + E_{sc}(\textbf{R},\mathcal{G})\,,
\label{Eqn:Electrostatics_FD} 
\end{align}
where the pseudocharge density $b(\br, \textbf{R}, \mathcal{G})$ admits the decomposition:
\begin{align}
\label{Eqn:Net_pseudocharge_FD}
b(\br, \textbf{R}, \mathcal{G}) = \sum_{\Upsilon_{\zeta,\mu} \in \mathcal{G}}\sum_{J=1}^{N} b_J(\br, \Upsilon_{\zeta,\mu} \circ \bR_J)\,.
\end{align}
Note that the term $\bar{E}_{sc}$ can similarly  be reduced to $E_{sc}$ by restricting the integrals over $\R^3$ to $\Omega$, not described here for brevity.

The fifth and final term in the energy functional (Eq.~\ref{Eqn:free_energy_per_FD}) represents the electronic entropy contribution to the free energy per fundamental domain. To obtain this term from Eq.~\ref{Eqn:Electronic_Entropy}, we include contribution from the occupations of $N_s$ electronic states from each diagonal block of the Hamiltonian to arrive at:
\begin{align}
\label{Eqn:Electronic_Entropy_FD}
S(\bg(\widehat{\mathcal{G}})) = -2k_B\sum_{n=1}^{N_{s}} \bigg(\frac{1}{\mathfrak{N}}\sum_{\nu=0}^{\mathfrak{N}-1} \fint \big( g_{n}(\nu,\eta) \log g_{n}(\nu,\eta) +  (1 - g_{n}(\nu,\eta)) \log (1 - g_{n}(\nu,\eta)) \big) \, \mathrm{d\eta}\bigg)\,.
\end{align}
\paragraph{Variational problem} The symmetry-adapted variational problem for determining the electronic ground state takes the form:
\begin{align}
\label{Eqn:free_energy_minimization_FD}
& \mathcal{F}_0(\bR,\mathcal{G}) =\min_{\Psi(\widehat{\mathcal{G}}), \bg(\widehat{\mathcal{G}})} \mathcal{F}\big(\Psi(\widehat{\mathcal{G}}) , \bg(\widehat{\mathcal{G}}), \textbf{R}, \mathcal{G}\big), \nonumber \\
 \text{s.t.} \int_{\Omega} \psi_{i}^*(\br,\nu,\eta) \psi_{j}(\br,\nu,\eta) \, \mathrm{d\br} = \delta_{i,j} & \,, \quad \text{for} \quad i,j=1,\ldots,N_s, \quad \nu \in \{0,\ldots,\mathfrak{N}-1\}, \quad {\eta} \in \left[-\frac{\pi}{H}, \frac{\pi}{H} \right]\,; \\
&\text{and}\; 2 \sum_{n=1}^{N_s} \bigg(\frac{1}{\mathfrak{N}}\sum_{\nu=0}^{\mathfrak{N}-1} \fint g_{n}(\nu,\eta) \, \mathrm{d\eta}\bigg) = N_e\,, \nonumber
\end{align}
where the symmetry-adapted orthogonality constraints follow from the orthogonality of the orbitals associated with the distinct characters of $\mathcal{G}$. The constraint on the number of electrons $N_e$ in the fundamental domain is obtained by including the contribution from the occupations of $N_s$ electronic states from each diagonal block of the Hamiltonian. 
\paragraph{Kohn-Sham equations} On taking variations of the constrained minimization problem presented above, we arrive at the following symmetry-adapted Kohn-Sham equations on the fundamental domain $\Omega$:
\begin{align}
\label{Eqn:FD_Hamiltonian_1}
\left( \mathcal{H}_{KS}\big(\Psi(\widehat{\mathcal{G}}) , \bg(\widehat{\mathcal{G}}), \textbf{R}, \mathcal{G}\big) \equiv -\frac{1}{2} \nabla^2 + V_{xc} + \phi +\mathcal{V}_{nl} \right) \psi_n(\br, \nu, \eta) &= \lambda_n(\nu, \eta)\,\psi_n(\br, \nu, \eta)\,, \\
\text{for} \quad n=1,2, \ldots, N_s \,, \quad \nu \in \{0,\ldots,\mathfrak{N}-1\}, & \quad {\eta} \in \left[-\frac{\pi}{H}, \frac{\pi}{H} \right] \,, \nonumber
\end{align}
subject to the boundary conditions given by Eqs.~\ref{Eqn:psi_theta_BC}-\ref{Eqn:psi_R_BC}. In the above equation, $V_{xc} = \displaystyle \frac{\delta E_{xc}}{\delta \rho}$ is the exchange-correlation potential, $\phi$ is the solution of the following Poisson problem on $\Omega$:
\begin{align}
\label{Eqn:poisson_problem_FD}
-\frac{1}{4 \pi} \nabla^2 \phi(\br, \textbf{R}, \mathcal{G}) = \rho(\br)+b(\br, \textbf{R}, \mathcal{G}) \,,
\end{align}
subject to the boundary conditions given by Eqs.~\ref{Eqn:phi_theta_BC}-\ref{Eqn:phi_R_BC}. The symmetry-adapted nonlocal operator $\mathcal{V}_{nl}$ acts on a function $f$ defined over the fundamental domain as:
\begin{align}
[\mathcal{V}_{nl} f ](\br) =\sum_{J=1}^{N} \sum_{p \in \mathcal{A}_{J}}\gamma_{J;p}\;\widetilde{\chi}_{J;p}(\bR_J, \br)\int_{\Omega}\widetilde{\chi}_{J;p}^{*}(\bR_J, \textbf{y}) f(\textbf{y})\, \mathrm{d\textbf{y}}\,.
\end{align}
In the above equations and those that follow, the electron density is calculated using Eq.~\ref{Eqn:Sym_electron_density_1}, and the occupations are determined through the Fermi-Dirac function, i.e.,
\begin{align}
\label{Eqn:occupation_Fermi_Dirac_FD}
g_{n}(\nu,\eta) = \bigg( 1 + \exp{\bigg( \frac{\lambda_n(\nu,\eta) - \lambda_F}{k_B T} \bigg)} \bigg)^{-1}\,, 
\end{align}
where the Fermi level $\lambda_F$ is determined by satisfying the constraint on the number of electrons in the fundamental domain (Eq.~\ref{Eqn:free_energy_minimization_FD}). 


\paragraph{Harris-Foulkes functional} 
Once the electronic ground state is determined by solving the above equations self-consistently, the free energy per fundamental domain can be computed through Eq.\ \ref{Eqn:free_energy_per_FD}, or alternately through the symmetry-adapted Harris-Foulkes  functional: 
\begin{align} 
\nonumber
\mathcal{F}_0(\bR,\mathcal{G}) & =  2 \sum_{n=1}^{N_s} \bigg(\frac{1}{\mathfrak{N}}\sum_{\nu=0}^{\mathfrak{N}-1} \fint \,g_{n}(\nu,\eta) \lambda_{n}(\nu,\eta) \, \mathrm{d\eta}\bigg) + E_{xc}(\rho)  - \int_{\Omega} V_{xc}(\rho(\br))\rho(\br)\, \mathrm{d\br}
\\ &+  \frac{1}{2} \int_{\Omega}\big(b(\br,\bR, \mathcal{G})-\rho(\br)\big) \phi(\br,\bR, \mathcal{G}) \, \mathrm{d\br} + E_{sc}(\bR, \mathcal{G}) -TS(\bg(\widehat{\mathcal{G}}))  \label{Eqn:Harris_Foulkes_FD}
\,.
\end{align}

\paragraph{Atomic forces} 
The symmetry-adapted Hellmann-Feynman atomic forces in the Cartesian coordinate system take the form:
\begin{align}
\nonumber
\textbf{f}_J &= - \frac{\partial \mathcal{F}_0(\bR,\mathcal{G})}{ \partial \textbf{R}_J} \\\nonumber
&=\sum_{\Upsilon_{\zeta, \mu} \in \mathcal{G}} \mathfrak{R}^{\mathfrak{N}-\zeta} \int_{\Omega} \nabla b_{J}(\br,\Upsilon_{\zeta, \mu} \circ \bR_{J}) \phi(\br,\bR,\mathcal{G})\, \mathrm{d\br} + \mathbf{f}_{sc,J}(\bR) \\\nonumber
&- 4 \sum_{n=1}^{N_s} \Bigg(\frac{1}{\mathfrak{N}}\sum_{\nu=0}^{\mathfrak{N}-1} \fint \,g_{n}(\nu,\eta) \sum_{p \in \mathcal{A}_J}\gamma_{J;p} \text{Re}\Bigg[\,\bigg(\int_{\Omega}\psi_n^{*}(\br,\nu,\eta)\,\widetilde{\chi}_{J;p}(\bR_J, \br,\nu,\eta)\,\mathrm{d\br}\bigg)\\
&\times \bigg( \sum_{\substack{\bR_{J'} =  \Upsilon_{\zeta, \mu} \circ \bR_{J}\\ \Upsilon_{\zeta, \mu} \in \mathcal{G}}}\!\mathfrak{R}^{\mathfrak{N}-\zeta}  \int_{\Omega}\nabla  \psi_n(\br,\nu,\eta){\chi}_{J;p}^*(\bR_{J'}, \br) e^{i\{ \nu(\theta_J-\theta_{J'})+ \eta (z_J - z_{J'})\}}\, \mathrm{d\br} \bigg)\Bigg]\Bigg) \mathrm{d\eta}\,,
\label{Eqn:Hellman_Feynman_FD} 
\end{align}
where $\nabla$ denotes the  Cartesian gradient operator and $\mathbf{f}_{sc,J}(\bR,\mathcal{G})= -\frac{\partial E_{sc}(\bR,\mathcal{G})}{\partial \bR_J}$. Note that in deriving the nonlocal component of the force, we have transferred the derivative on the projectors (with respect to atomic position) onto the orbitals (with respect to atomic position).  This is because the orbitals are typically smoother than the projector functions, therefore resulting in substantially more accurate atomic forces \citep{hirose2005first, Ghosh2017cluster, Ghosh2017extended}.

\subsubsection{Time reversal symmetry}
In the absence of magnetic fields, the symmetry-adapted formulation presented above can be further reduced by employing time-reversal symmetry\citep{Martin2004, Banerjee_Elliott_Symmetry_paper}. Specifically, for $\nu \in \{1,2,\ldots,\mathfrak{N}-1\}$, we have the relations:
\begin{align}
& \lambda_n(\nu,\eta) = \lambda_n(\mathfrak{N}-\nu,-\eta) \,, \quad \psi_n(\br, \nu,\eta) = \psi_n^*(\br, \mathfrak{N}-\nu,-\eta) \,, 
\end{align}
and for $\nu=0$:
\begin{align}
\lambda_n(0,\eta) = \lambda_n(0,-\eta) \,, \quad \psi_n(0,\eta) = \psi_n^*(0,-\eta) \,.
\end{align}
As a result, the character space $(\nu,\eta)$ that needs to be considered for the Kohn-Sham equations as well as the calculation of the free energy and atomic forces is essentially halved. 

\subsubsection{Computational cost reduction due to cyclic symmetry adaptation} 
The proposed formulation is able to achieve significant reduction in the computational cost of DFT simulations for systems possessing cyclic symmetry.  Consider the nonlinear eigenvalue problem, which forms the dominant part of Kohn-Sham calculations and scales cubically with the problem size asymptotically. Since the problems associated with different $\nu$ are independent, the cyclic symmetry adaptation  translates to a factor of $\mathcal{O}(\mathfrak{N}^2)$ reduction in computational cost. This can result in tremendous savings, particularly for systems where $\mathfrak{N}$ is large, such as those studied in this work. Reductions in cost also extend to the solution of the Poisson equation as well as the calculation of the energy and forces, where the factor is the more modest $\mathcal{O}(\mathfrak{N})$. Note that even further reductions are achieved in practice while solving for the electronic ground state using the self-consistent field (SCF) method, due to the notable reduction in symmetry breaking and charge sloshing type instabilities \cite{Banerjee2016cyclic}. In addition, the implementations are more amenable to scalable parallel computations, due to the significantly fewer global communications. Indeed, such communications are significantly reduced because the orbitals associated with distinct characters are automatically orthogonal to each other.

\section{Numerical Implementation} \label{Section:Implementation}
We implement the proposed formulation within the real-space finite-difference DFT code SPARC \cite{Ghosh2017cluster,Ghosh2017extended}. Due to the nature of SPARC's implementation, we employ the Bloch-type ansatz for convenience:
\begin{align}
\psi_{n}(\br,\nu,\eta) = e^{-i(\nu \theta+\eta z)} u_{n}(\br,\nu,\eta) \,,
\label{Eqn:Bloch_theorem_3}
\end{align}
where $u_{n}(\br,\nu,\eta)$ is group invariant, i.e., for any $\Upsilon_{\zeta,\mu}  \in \mathcal{G}$, 
\begin{align}
u_n(\Upsilon_{\zeta,\mu}\circ\br,\nu,\eta) = u_n(\br,\nu,\eta) \,.
\end{align}
As a result, $u_n$ replaces $\psi_n$ as the primary unknown functions that are being solved for in the symmetry-adapted Kohn-Sham equations, the modified form of which can easily be derived using the above ansatz, and are  therefore not reproduced here for the sake of brevity.  

The geometry of $\Omega$ motivates the use of cylindrical polar coordinates, wherein the Laplacian and the Cartesian gradient operator take the form:
\begin{align}
& \nabla^2 \equiv \bigg(\frac{\partial^2}{\partial r^2} + \frac{1}{r} \frac{\partial}{\partial r} + \frac{1}{r^2}\frac{\partial}{\partial \theta} + \frac{\partial^2}{\partial z^2} \bigg) \,,\label{Eqn:Lap:Polar}\\
& \nabla \equiv \left(\cos \theta \frac{\partial}{\partial r} - \frac{\sin \theta}{r} \frac{\partial}{\partial \theta} \right)\mathbf{e}_x + \left(\sin \theta \frac{\partial}{\partial r} + \frac{\cos \theta}{r} \frac{\partial}{\partial \theta} \right)\mathbf{e}_y + \frac{\partial}{\partial z} \mathbf{e}_z\,.
\label{Eqn:nabla_Cartesian_cylindrical}
\end{align} 
The fundamental domain $\Omega$ is discretized using a finite-difference grid with spacing $h_r$, $h_{\theta}$, and $h_z$ along the $r$, $\theta$, and $z$ directions, respectively. This implies that $R_2-R_1=n_r h_r$, $\Theta=n_{\theta}h_{\theta}$ and $H=n_z h_z$, for natural numbers $n_r,n_{\theta}$ and $n_z$. Each finite-difference node is indexed using a triplet of the form $(i,j,k)$, with $i=1,2,\ldots, n_r$, $j=1,2,\ldots, n_{\theta}$, and $k=1,2,\ldots, n_z$. Using the central finite-difference approximation, we approximate the partial first derivatives as:
\begin{eqnarray}\label{Eqn:gradient:approximate}
\frac{\partial f}{\partial r} \bigg|^{(i,j,k)} &\approx&  \sum_{p=1}^{n_o} \bigg( \tilde{w}_{p,r}  ( f^{(i+p,j,k)} - f^{(i-p,j,k)}) \bigg) \,,  \nonumber \\
\frac{\partial f}{\partial \theta} \bigg|^{(i,j,k)} &\approx&  \sum_{p=1}^{n_o} \bigg(  \tilde{w}_{p,\theta} ( f^{(i,j+p,k)} - f^{(i,j-p,k)}) \bigg) \,, \nonumber \\
\frac{\partial f}{\partial z} \bigg|^{(i,j,k)} &\approx& \sum_{p=1}^{n_o} \bigg(\tilde{w}_{p,z} ( f^{(i,j,k+p)} - f^{(i,j,k-p)}) \bigg) \,,
\end{eqnarray}
and similarly the partial second derivatives as:
\begin{eqnarray}\label{Eqn:SecondDerivative:approximate}
\frac{\partial^2 f}{\partial r^2}  \bigg|^{(i,j,k)} &\approx& \sum_{p=0}^{n_o} \bigg( w_{p,r} (f^{(i+p,j,k)} + f^{(i-p,j,k)} ) \bigg) \,, \nonumber \\ 
\frac{\partial^2 f}{\partial {\theta}^2} \bigg|^{(i,j,k)} &\approx& \sum_{p=0}^{n_o} \bigg( w_{p,\theta} ( f^{(i,j+p,k)} + f^{(i,j-p,k)} \bigg)  \,,
\nonumber \\ 
\frac{\partial^2 f}{\partial z^2} \bigg|^{(i,j,k)} &\approx&   \sum_{p=0}^{n_o} \bigg( w_{p,z}( f^{(i,j,k+p)} + f^{(i,j,k-p)} ) \bigg) \,,
\end{eqnarray}
with $f^{(i,j,k)}$ representing the value of the function $f$ at the node $(i,j,k)$. Denoting $s \in \{r, \theta, z\}$, the weights that appear in the above expressions can be written as \citep{mazziotti1999spectral,Suryanarayana2014524}:
\begin{eqnarray}
w_{0,s} & = & - \frac{1}{h_s^2} \sum_{q=1}^{n_o} \frac{1}{q^2} \,, \,\,
\nonumber \\
w_{p,s} & = & \frac{2 (-1)^{p+1}}{h_s^2 p^2} \frac{(n_o!)^2}{(n_o-p)! (n_o+p)!} \,, \,\, p=1, 2, \ldots, n_o\,, \label{Eqn:weightssecondderivative} \\
\tilde{w}_{p,s} & = & \frac{(-1)^{p+1}}{h_s p} \frac{(n_o!)^2}{(n_o-p)! (n_o+p)!} \,, \,\, p=1, 2, \ldots, n_o\,.
\end{eqnarray}
Due to the curvilinear nature of the underlying coordinate system, the Laplacian and Hamiltonian matrices resulting from the above discretization scheme are non-Hermitian, even though the infinite-dimensional operators from which they arise are Hermitian\citep{Banerjee2016cyclic, gygi1995real}. For a fixed finite-difference order however, as the discretization is refined, the discrete Laplacian and Hamiltonian matrices approach Hermitian matrices and the eigenvalues of these matrices turn out to be either real, or they have vanishingly small imaginary parts\cite{gygi1995real, Banerjee2016cyclic}. Hence this issue does not negatively impact the physical results obtained or their implications.

We approximate the integrals over the fundamental domain by employing the following quadrature rule:
\begin{equation} \label{Eqn:IntApprox}
\int_{\Omega} f(\br) \, \mathrm{d\br} \approx  h_r h_{\theta} h_z \sum_{i=1}^{n_r} \sum_{j=1}^{n_\theta} \sum_{k=1}^{n_z} r_i f^{(i,j,k)}\,,
\end{equation} 
with $r_i$ denoting the radial coordinate of the finite-difference node indexed by $(i,j,k)$. We enforce periodic boundary conditions by mapping any index that does not correspond to a node in the finite-difference grid to its periodic image within $\Omega$. We enforce zero Dirichlet boundary conditions by setting $f^{(i,j,k)}=0$ for any index that does not correspond to a node in the finite-difference grid.
Following the strategy proposed previously\cite{Phanish2012,Ghosh2017cluster}, we use the discrete Laplacian to directly compute the pseudocharges from the local parts of the pseudopotentials, while assigning them to the grid. Due to the presence of translational symmetry along $\textbf{e}_z$, evaluating integrals such as Eq.~\ref{Eqn:Sym_electron_density_1} requires us to discretize the domain of the variable $\eta$, allowing for a discrete representation of the complex characters associated with the periodic symmetry. Accordingly, we utilize the Monkhorst-Pack \cite{monkhorst1976special} grid for sampling the interval $[-\frac{\pi}{H},\frac{\pi}{H}]$ and approximate the averaged integral of any function over the interval as:
\begin{equation} \label{Eqn:BZ:Integration}
\fint f(\eta) \, \mathrm{d \eta} \approx \sum_{b=1}^{N_{\eta}} w_b f(\eta_b)\,.
\end{equation}
Here $\eta_b$ and $w_b$ denote the integration nodes and weights, respectively.  The total number of discretized characters in the computation (i.e., ``\textbf{k}-points'' in the language of periodic DFT calculations) is denoted as $N_K \approx \frac{1}{2} (\mathfrak{N} \times N_{\eta})$, where time-reversal symmetry has been used to reduce the total number by a factor of two approximately.

We use the Chebyshev polynomial filtered subspace iteration (CheFSI) technique \cite{zhou2006self,zhou2006parallel} in conjunction with potential mixing for computing the electronic ground state  corresponding to Eq.~\ref{Eqn:FD_Hamiltonian_1}. Within the CheFSI method, we employ Arnoldi iterations\citep{saad2003iterative} for calculating the extremal eigenvalues of the Hamiltonian, and LAPACK\cite{laug} for solving the projected subspace eigenproblem. We solve the linear system corresponding to the Poisson problem (Eq.~\ref{Eqn:poisson_problem_FD}) using the block-Jacobi preconditioned \cite{golub2012matrix} Generalized minimal residual method (GMRES) \cite{saad1986gmres}. We calculate the Fermi energy using Brent's method \cite{press2007numerical}, and use Periodic Pulay extrapolation \cite{Banerjee2016PeriodicPulay} for accelerating convergence of the SCF iterations. We calculate the ground state free energy using the symmetry-adapted Harris-Foulkes type functional (Eq.~\ref{Eqn:Harris_Foulkes_FD}). If and when required, we employ the FIRE algorithm \cite{bitzek2006structural} for performing structural relaxations.

The real-space discretization naturally lends itself to parallelization via domain decomposition.  In addition, the eigenvalue problems corresponding to distinct characters can be solved independently. Therefore, we employ two levels of parallelization: the first being over the different values of the characters and the second being over the spatial domain. The former is achieved by uniformly distributing the list of $N_K$ eigenvalue problems among $N^{proc}_K$ processors ($N^{proc}_K \leq N_K$). The latter is achieved by partitioning the fundamental domain $\Omega$ among $N^{proc}_D$ processors as:
\begin{align}
\Omega = \bigcup \limits_{p=1}^{N^{proc}_D} \Omega_p,
\end{align} 
and assigning the portion of the calculation associated with the partition $\Omega_p$ to the $p^{th}$ processor.  The total number of processors employed in this two level parallelization scheme is therefore $N^{proc} = N^{proc}_D \times N^{proc}_K$. We use the Portable, Extensible Toolkit for Scientific computations (PETSc)\citep{balay2018petsc} suite of data structures and routines, in conjunction with the Message Passsing Interface (MPI)\citep{gropp1999using} for implementation and parallelization of our computational routines.

\section{Results and discussion} \label{Section:ExamplesResults}
The main objects of study in this work are single walled nanotubes of carbon, silicon, germanium, and tin, collectively referred to here as X (X=C,Si,Ge,Sn) nanotubes. These 1D nanostructures are formed by rolling their 2D sheet counterparts: graphene, silicene, germanene and stanene, collectively referred to here as Xenes. Depending on whether the direction of rolling is armchair or zigzag, the nanotubes can be classified as  armchair or zigzag, respectively. Since zigzag carbon nanotubes have distinct electronic properties based on their radius \citep{saito1998physical}, we further classify the zigzag X nanotubes as type I, II, or III, depending on whether $\text{mod}(\mathfrak{N},3)=1, 2$ or $0$. Both X nanotubes and Xene sheets are known to demonstrate unusual and fascinating material properties \cite{Martel1998,javey2003ballistic,popov2004carbon,gong2009nitrogen,park2009silicon,wu2012stable,park2011germanium,li2011controlled,zhao2006porous,xu2013graphene,bhimanapati2015recent,butler2013progress,naguib201425th,fiori2014electronics,koppens2014photodetectors},  motivating their choice here as well as in a number of previous electronic structure studies \cite{blase1994hybridization,spataru2004excitonic,yang2000electronic,fagan2000ab,benedict1995static,zhang2003silicon,yang2005electronic,giovannetti2008doping,vogt2012silicene,davila2014germanene,zhu2015epitaxial}. 

The calculations here utilize the LDA \citep{Kohn1965} to model the exchange-correlation functional, with the Perdew-Wang parametrization \cite{perdew1992accurate} of the correlation energy as calculated by Ceperley and Alder \cite{Ceperley1980}; smearing of  $k_B T = 0.001$ Ha, treated here as a numerical parameter to aid SCF convergence rather than the actual temperature; and Troullier-Martins norm conserving pseudopotentials\citep{Troullier}. With these choices, the ground state interatomic distance ($a$), and out of plane buckling distance ($\delta$) for the planar Xene sheets are as reported in Table~\ref{Table:LatticeParameters}. The agreement of these values with the literature is generally quite good, thus giving us confidence in the quality of the simulations. 

\begin{table}[htb]
\centering
\begin{tabular}{c  c  c}
\hline \vspace{-3mm} \\
Xene & $a$ (\angstrom) & $\delta$ (\angstrom)  \\   
\hline
C  & 1.407 (1.408\cite{kerszberg2015})&  - \\
Si  &  2.200 (2.207\cite{Buda}) & 0.404 (0.437\cite{Buda}) \\
Ge & 2.232 (2.290\cite{Buda})& 0.566 (0.647\cite{Buda})\\
Sn  & 2.522 (2.611 \cite{zhou2016quantum})& 0.699 (0.822 \cite{zhou2016quantum})\\
\hline
\end{tabular}
\caption{Equilibrium lattice parameters for the Xene sheets. The numbers in parenthesis are values from literature.}
\label{Table:LatticeParameters}
\end{table}

\begin{figure}[htb]
\includegraphics[width=0.6\textwidth]{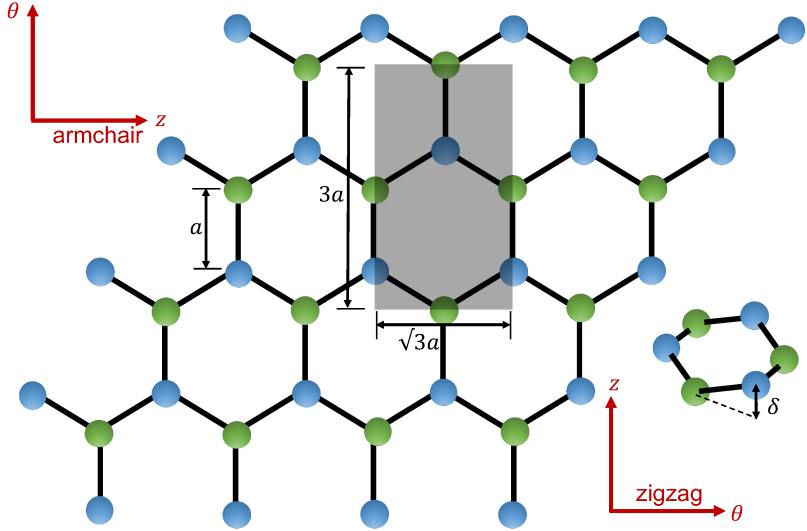}
\caption{Schematic showing the geometry of Xene (X=C,Si,Ge,Sn) sheets, orthogonal unit cell of the Xene sheet rolled to generate the structure within the fundamental domain of the X nanotube, and orientation of the resulting armchair and zigzag X nanotubes.} 
\label{Fig:Material_Geometry}
\end{figure}

The combined use of cyclic and periodic symmetries allows X nanotubes to be represented by just 4 atoms within the fundamental domain, i.e., the fundamental domain corresponds to the rolling of the  4-atom orthogonal unit cell in the Xene sheet, as shown in Fig.~\ref{Fig:Material_Geometry}. The angle formed by the fundamental domain $\Theta = 2\pi/\mathfrak{N}$, where $\mathfrak{N}$ depends on the the radius of the nanotube and the interatomic distance in the flat sheets. Specifically, in the absence of relaxation effects \footnote{X (X=C, Si, Ge, Sn) nanotubes formed by rolling the corresponding Xene sheets are not necessarily at the structural ground state \cite{Sanchez1999}. Indeed, for small radii nanotubes, the atoms can experience atomic forces as large as $0.01$ Ha/Bohr. However, for the large radii nanotubes studied in this work, the maximum component of the atomic force in the initial configuration is less than $3\times 10^{-3}$ Ha/Bohr, with the value becoming smaller as the radius gets larger. Therefore, the adopted procedure provides a very good guess for the atomic positions. Moreover, since there is no noticeable change in the results between this configuration and the structural ground state,  we do not perform any structural optimization steps in this work. \label{Footnote:Forces}}, $\mathfrak{N} = \pi/\sin^{-1}\left(\frac{L}{2R}\right)$, where $L = 3a$ and $\sqrt[]{3}a$ for armchair and zigzag nanotubes, respectively. The corresponding heights of the fundamental domain are $H = \sqrt[]{3}a$ and $3a$, respectively. The radii $R_{in}$ and $R_{out}$ are chosen such that all atoms are at least $11$ Bohr away from the boundaries in the radial direction, so as to allow sufficient decay of the electron density and orbitals.

We discretize the governing equations using a twelfth-order accurate finite-difference discretization. Subsequent to a convergence analysis with respect to mesh size $h = \displaystyle \max \left\lbrace h_r, \bigg(\frac{R_{in}+R_{out}}{2}\bigg)h_{\theta}, h_z \right\rbrace$ (similar to Section~\ref{subsec:accuracy}) as well as an increasingly finer sampling of the values of $\eta$, the $h$ and $N_{\eta}$ listed in Table \ref{Table:DiscretzationParameters} are chosen for the X nanotube simulations in Sections~\ref{subsec:nanotube_electronic_studies} and \ref{subsec:sheet_bending_studies}, from which the physical properties of interest, namely bandgap variation with radius of X nanotubes and bending moduli of Xene sheets, are calculated. The chosen parameters ensure that the energy and atomic forces are converged to within $10^{-5}$ Ha/atom and $10^{-5}$ Ha/Bohr, respectively. Note that such high precision---significantly more stringent than that typically employed in DFT calculations---is essential to capture the small bandgap and energy variations that occur with respect to the radius.

\begin{table}[htb]
\centering
\begin{tabular}{c c  c}
\hline
X & $\quad h$ (Bohr) & $\quad N_{\eta}$\\
\hline
C & 0.125  & Armchair: $21$, Zigzag:  $13$   \\ 
Si & 0.250  & Armchair: $15$, Zigzag:  $9$  \\ 
Ge & 0.200  & {Armchair: $15$, Zigzag:  $9$  } \\ 
Sn & 0.250  & {Armchair: $15$, Zigzag:  $9$  } \\ 
\hline 
\end{tabular}
\caption{Real- and $\eta$-space discretization parameters for the X nanotube simulations in Sections~\ref{subsec:nanotube_electronic_studies} and \ref{subsec:sheet_bending_studies}, from which the physical properties of interest, i.e., variation of bandgap with radius of X nanotubes and the bending moduli of Xene sheets, are calculated.}
\label{Table:DiscretzationParameters}
\end{table} 

\subsection{Convergence and accuracy} \label{subsec:accuracy}
\begin{figure}[htb]\centering
\subfloat[Energy]{\includegraphics[keepaspectratio=true,width=0.45\textwidth]{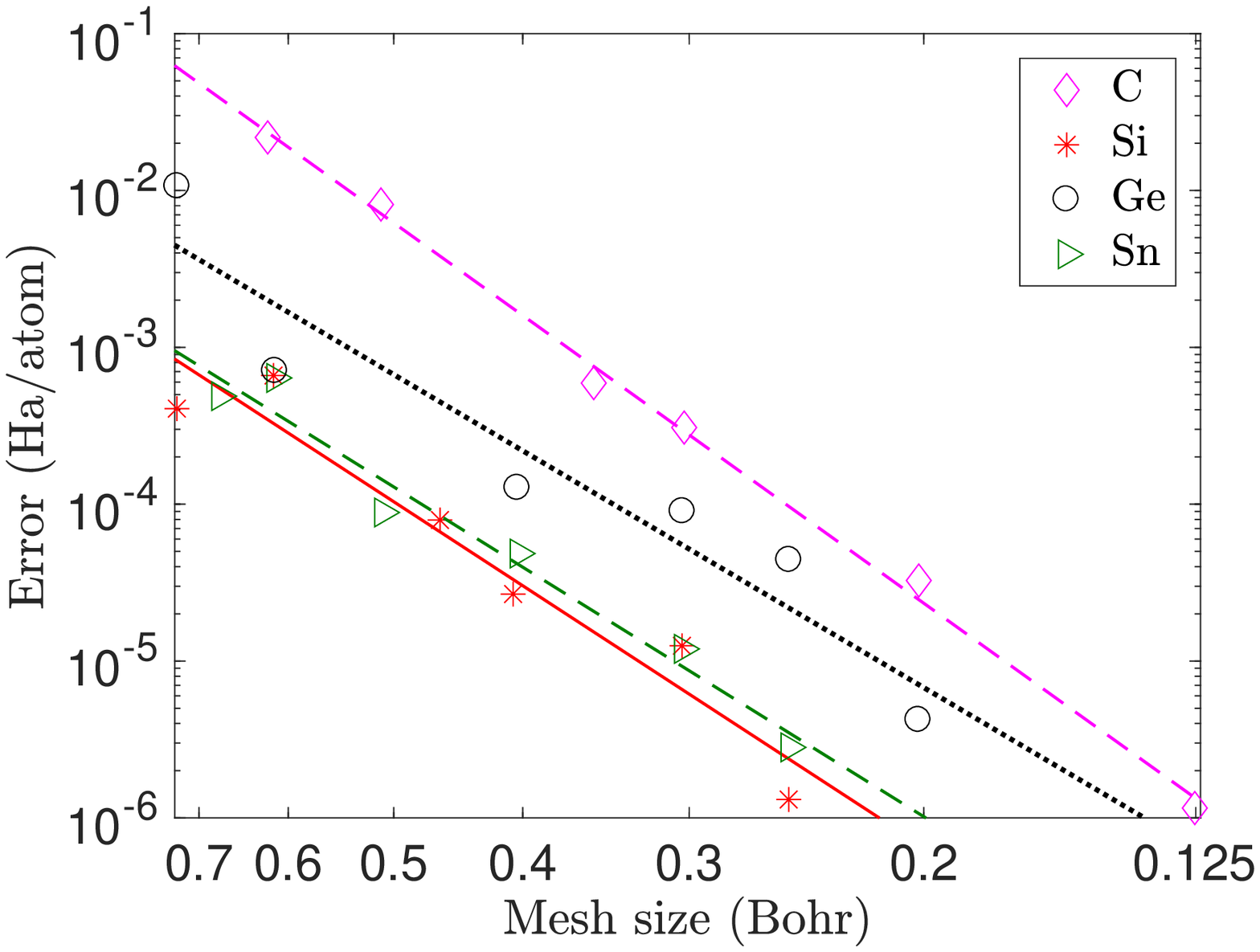}\label{Fig:EnergyMesh}}
\subfloat[Atomic forces]{\includegraphics[keepaspectratio=true,width=0.45\textwidth]{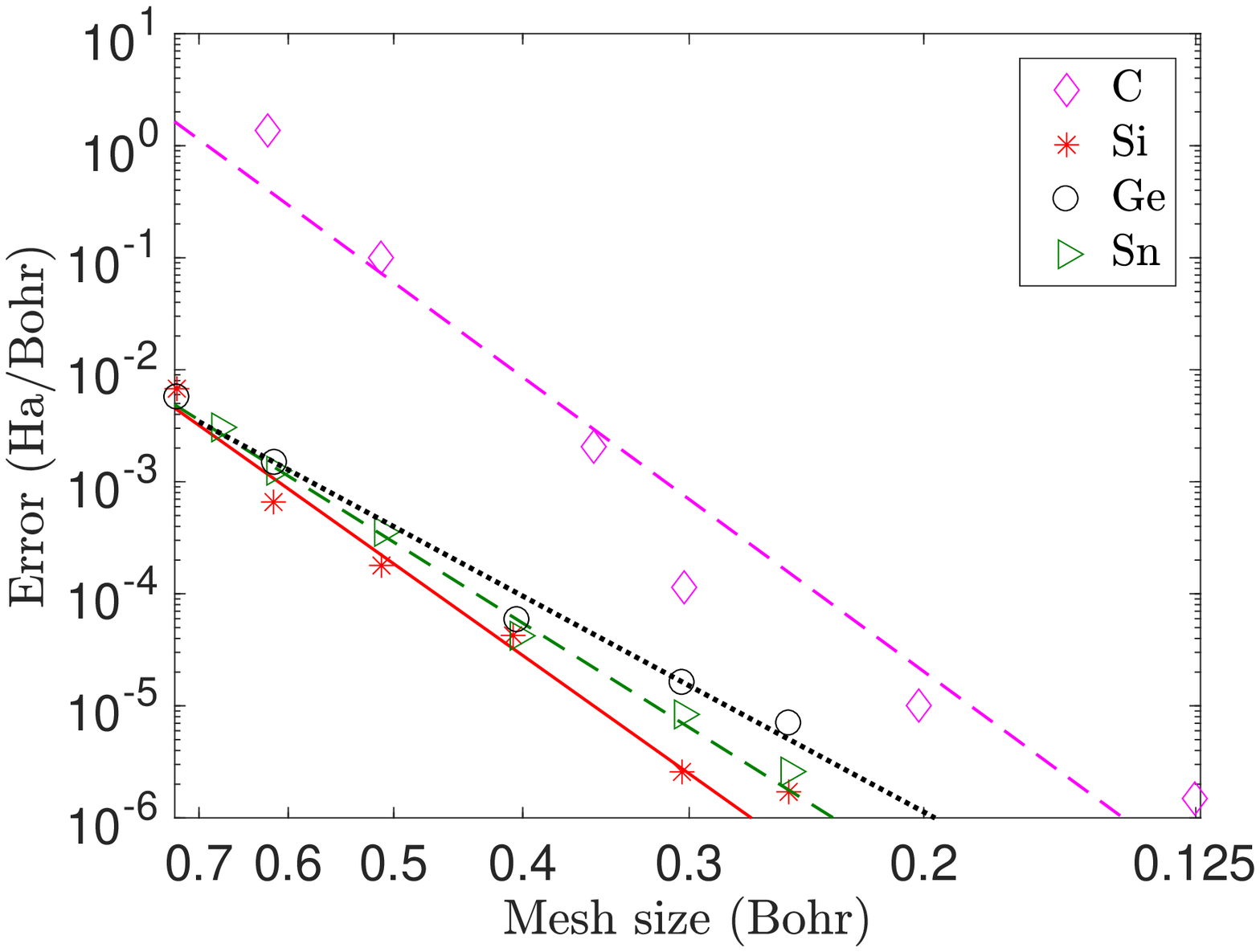}\label{Fig:ForceMesh}}
\caption{Convergence in energy and atomic forces of X (X=C,Si,Ge,Sn) nanotubes as a function of mesh size $h$. The error in the energy is defined to be the magnitude of the difference, and the error in the forces is defined to be the maximum (in magnitude) difference  in any component. The straight lines represent fits to the data.} 
\label{Fig:Mesh_Convergence}
 \end{figure}

To assess the accuracy of the proposed formulation and implementation, we first verify the convergence of the energy as well as the atomic forces with respect to spatial discretization, i.e., mesh size $h$. As representative systems, we choose zigzag X (X=C,Si,Ge,Sn) nanotubes with radii $0.90$, $0.98$, $0.99$, and $1.13$ nm, corresponding to $\mathfrak{N}=23$, $16$, $16$, and $16$, respectively.  Note that the radii of these tubes is sufficiently small for the atoms to experience significant atomic forces. Without loss of generality, only the $\eta = 0$ point  is included for this numerical test (equivalent to a $\Gamma$-point calculation in traditional DFT). It is clear from the results in Fig.~\ref{Fig:Mesh_Convergence} that there is systematic convergence of both the energy and atomic forces to reference values obtained for $h = 0.1$ Bohr.  On fitting the data, we find average convergence rates in the energy and atomic forces of $5.5$ and $7.8$, respectively, comparable to those obtained by the analogous real-space formalism for affine coordinate systems  \cite{Ghosh2017cluster,Ghosh2017extended,sharma2018real}. 

In order to further verify the accuracy of the proposed method, we compare the results at $h = 0.1$ Bohr with highly converged values obtained by the established planewave code ABINIT \citep{gonze2009abinit,ABINIT}. We find that there is agreement to within $6 \times 10^{-5}$ Ha/atom and $1 \times 10^{-4}$ Ha/Bohr in the energy and atomic forces, respectively. Indeed, even better agreement would have been possible, but for the significant challenge in converging ABINIT  to finer levels of accuracy. This is mainly due to the stagnation in the results with respect to vacuum \cite{Ghosh2017extended}, likely due to the inaccurate electrostatics resulting from the requirement of periodic boundary conditions. To confirm this, we have also compared with the real-space DFT code SPARC \cite{Ghosh2017cluster,Ghosh2017extended}, which is not only more efficient, but also does not suffer from the aforementioned stagnation. We have found that  there is indeed better agreement with SPARC, with energy and atomic forces differing by not more than $2 \times 10^{-6}$ Ha/atom and $1 \times 10^{-5}$ Ha/Bohr, respectively. 

Even though not demonstrated here, we have verified that the proposed method inherits a number of attractive features of the underlying SPARC framework. In particular, there is exponential convergence in the properties of interest with respect to the amount of vacuum in the radial direction. In addition, the computed energy and atomic forces are consistent, which allows for accurate geometry optimization and molecular dynamics simulations. Finally, the eggbox effect resulting from breaking of the translational and cyclic symmetry of the system is negligible, particularly at the mesh sizes considered in this work. Since these features have been demonstrated and discussed in detail previously for SPARC \cite{Ghosh2017cluster,Ghosh2017extended}, we do not repeat them here for the sake of brevity.


\subsection{Band structure of X (X=C, Si, Ge, Sn) nanotubes} \label{subsec:nanotube_electronic_studies}
We now use the proposed method to study the band structure of X (X=C, Si, Ge, Sn) nanotubes, with radii ranging from $R=1.3$ nm ($\mathfrak{N} = 23$) to $R=4.8$ nm ($\mathfrak{N} = 108$). Studies in literature have shown that larger radii carbon nanotubes are more prone to instability (e.g., flattening or collapse) when subject to hydrostatic stresses. \cite{chopra1995fully,gao1998energetics,elliott2004collapse,tangney2005structural,tang2005collapse,gadagkar2006collapse}. There is however substantial disagreement in the theoretically/computationally predicted values, with critical radii ranging from $1$ to $3.5$ nm at atmospheric conditions. Concerns regarding the validity of these predictions remain, particularly with the synthesis of carbon nanotubes having radii as large as $6$ nm \cite{cheung2002diameter}. The systems chosen here are motivated by the fact that large radii nanotubes are far less studied, particularly in the context of ab-initio calculations, where the associated computational cost is large. Moreover, such radii are required for accurately calculating the bending moduli of the Xene sheets, as done in Section~\ref{subsec:sheet_bending_studies}.

\begin{figure}[htb]\centering
\subfloat[Carbon: Variation along $\nu=0$]{\includegraphics[keepaspectratio=true,width=0.45\textwidth]{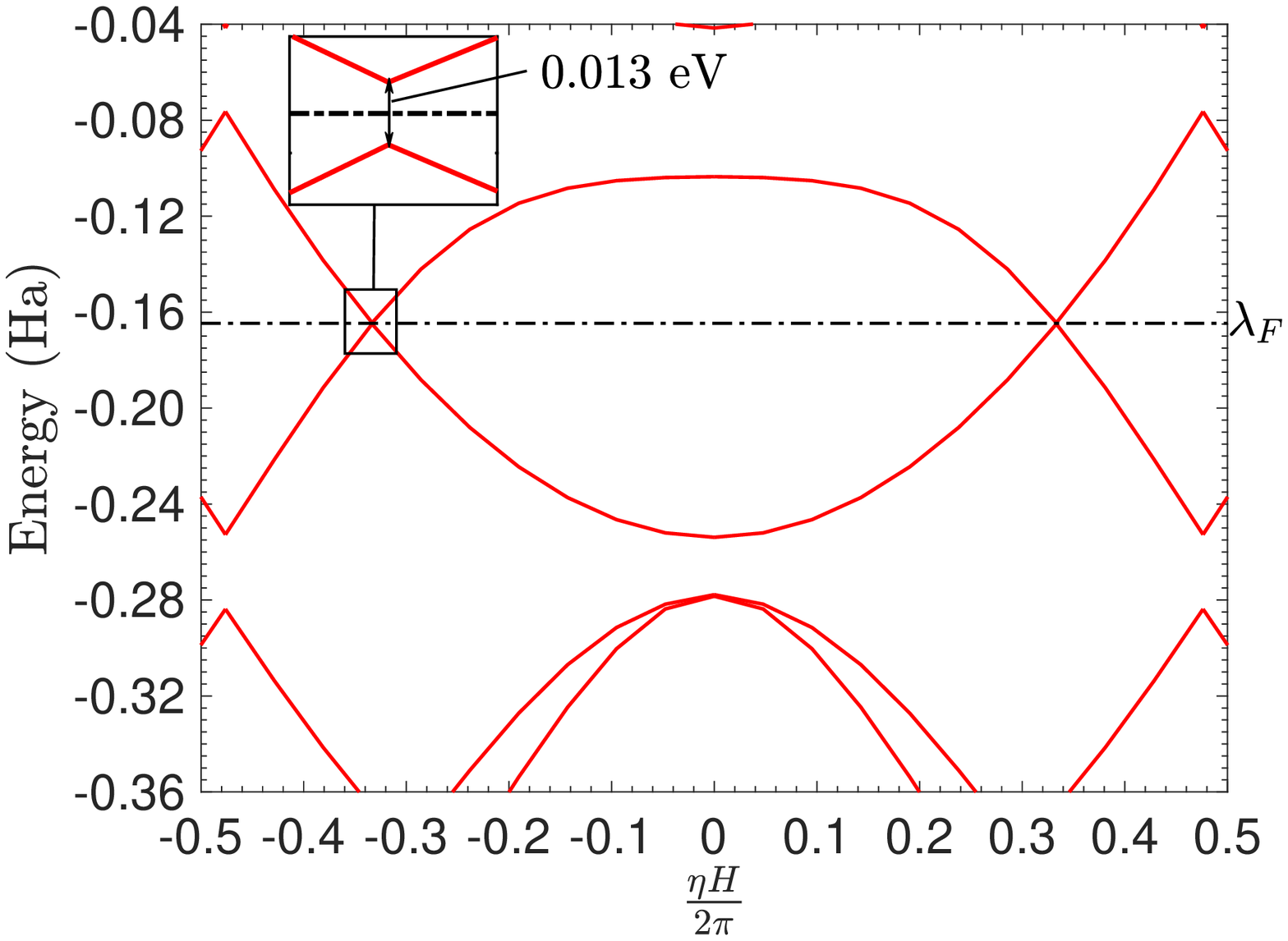}\label{Fig:GrapheneArmchairEta}}
\subfloat[Carbon: Variation along $\frac{\eta H}{2 \pi} = \frac{1}{3}$]{\includegraphics[keepaspectratio=true,width=0.45\textwidth]{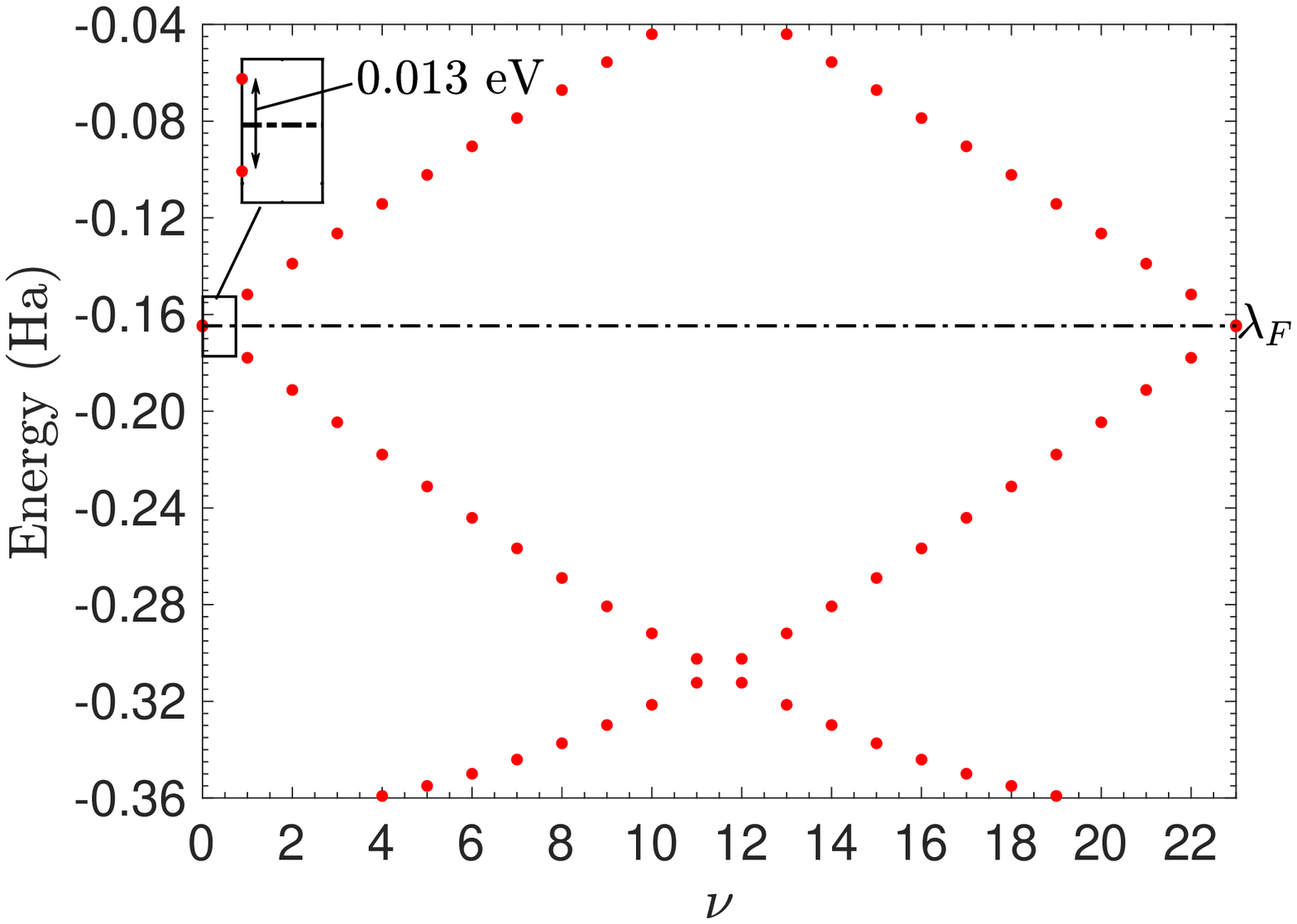}\label{Fig:GrapheneArmchairNu}}\\
\subfloat[Tin: Variation along $\nu=8$]
{\includegraphics[keepaspectratio=true,width=0.45\textwidth]{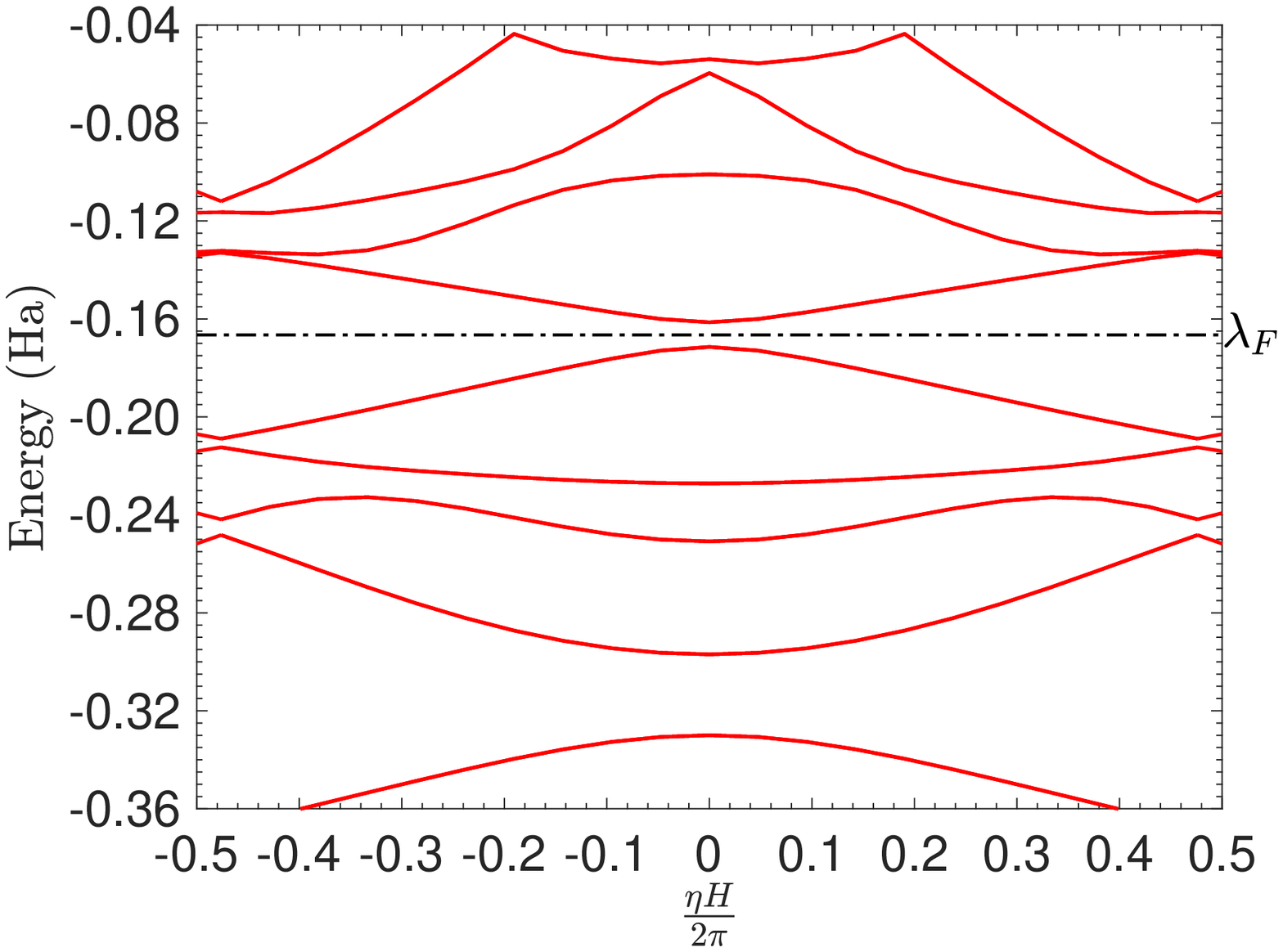}\label{Fig:StaneneArmchairEta}}
\subfloat[Tin: Variation along $\frac{\eta H}{2 \pi} = 0$]{\includegraphics[keepaspectratio=true,width=0.45\textwidth]{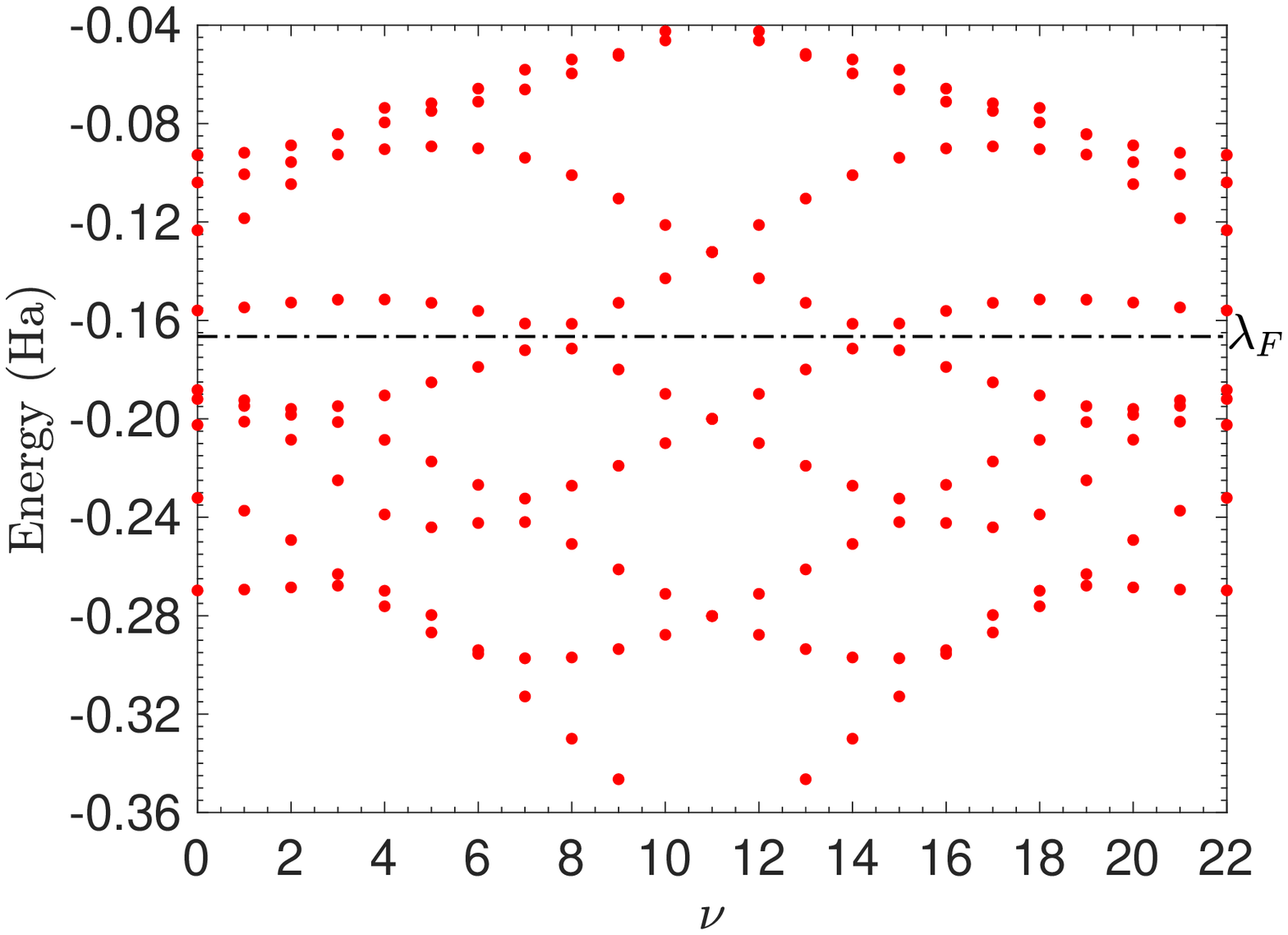}\label{Fig:StaneneArmchairNu}}\\
\caption{Band structure diagrams along specific line segments in $(\nu,\eta)$ space for carbon armchair and tin zigzag type I nanotubes of radii $R = 1.6$ nm ($\mathfrak{N} = 23$) and $R = 1.5$ nm ($\mathfrak{N} = 22$), respectively.}
\label{Fig:Bandstructure}
\end{figure}

We start by computing the symmetry-adapted band structure data for the aforementioned nanotubes in the discrete $(\nu,\eta)$ space. This ability to calculate and plot the variation of the eigenvalues with respect to both the character labels $(\nu,\eta)$, rather than with respect to $\eta$ alone (as is typically done), is a distinctive feature of the method  developed here. Such symmetry-adapted band structure diagrams have the advantage that they allow for significantly easier interpretation of the results, particularly for systems with complex band structure. See Fig.~\ref{Fig:Bandstructure} for representative band structure diagrams of carbon armchair ($R = 1.6$ nm, $\mathfrak{N} = 23$) and tin zigzag type I ($R = 1.5$ nm, $\mathfrak{N} = 22$) nanotubes along specific line segments in $(\eta, \nu)$ space. Note that for the band structure diagrams at fixed $\eta$ (i.e., Figs.~\ref{Fig:GrapheneArmchairNu} and \ref{Fig:StaneneArmchairNu}), $\nu$ can only take integer values. 

The calculated band structure data can be used to deduce whether the systems are metallic, semi-metallic or insulating. For insulating systems, the size of the bandgap can be determined by calculating the difference between the smallest eigenvalue above the Fermi level and largest eigenvalue below the Fermi level in all of $(\eta,\nu)$ space. We have found that all the X nanotubes are semiconducting. Specifically, there is a direct bandgap for the armchair nanotubes at $\left( \frac{\eta H}{2 \pi}, \nu \right) = \left( \frac{1}{3}, 0 \right)$ (or equivalently $\left( -\frac{1}{3}, 0 \right)$). In addition, apart from the zigzag type I carbon nanotube which has a direct bandgap at $\left( \frac{\eta H}{2 \pi}, \nu \right) = \left( 0, \frac{\mathfrak{N}-1}{3} \right)$ (or equivalently $\left( 0,\frac{2 \mathfrak{N}+1}{3} \right)$), the other zigzag nanotubes have a direct bandgap at $\left( \frac{\eta H}{2 \pi}, \nu \right) = \left( 0, \frac{\mathfrak{N}+2}{3} \right)$ (or equivalently $\left( 0, \frac{2 \mathfrak{N}-2}{3} \right)$), $\left( \frac{\eta H}{2 \pi}, \nu \right) = \left( 0, \frac{\mathfrak{N}+1}{3} \right)$ (or equivalently $\left( 0, \frac{2 \mathfrak{N}-1}{3} \right)$), and $\left( \frac{\eta H}{2 \pi}, \nu \right) = \left( 0, \frac{\mathfrak{N}}{3} \right)$ (or equivalently $\left( 0, \frac{2 \mathfrak{N}}{3} \right)$) for the type I, II, and III variants, respectively. \footnote{The implication (if any) of the bandgap being at different values of $\nu$ is not evident at this point, and requires a further in-depth study.}

The above results are generally in good agreement with those found in literature \cite{hamada1992new,saito1992electronic,mintmire1993properties,wang2017band,yang2005electronic}, apart from a few discrepancies. First, the bandgap in armchair germanium and tin nanotubes has previously been found to be indirect \cite{wang2017band}, whereas we predict a direct bandgap. This is likely due to the much larger nanotubes studied here. Second, while some electronic structure studies  \cite{wang2017band}, including the one here, have shown armchair silicon nanotubes to be semiconducting, others have found them to be metallic \cite{fagan2000ab,yang2005electronic}. A possible source of the disagreement is that  rather small planewave cutoffs have been used in cases where metallic behavior has been predicted. Finally, armchair carbon nanotubes are generally found to be metallic \cite{hamada1992new,saito1992electronic,mintmire1993properties}, however here we have obtained a nonzero (but vanishingly small) bandgap.  We have verified that the above disagreements are not an artifact of the proposed formulation or the chosen pseudopotential/exchange-correlation functional. For example, consider the armchair carbon nanotube of radius $R=1.55$ nm. Highly accurate planewave calculations using ABINIT predicts a bandgap identical to the value computed here (i.e., 0.0136 eV). On changing the pseudopotential from Troullier-Martins to ONCV \cite{hamann2013optimized}, the bandgap remains with a nearly identical value of 0.0133 eV, and on further changing the exchange-correlation functional from LDA to GGA \cite{perdew1996generalized}, the bandgap persists with a nearly identical value of 0.0130 eV. Since it is known from symmetry arguments that these nanotubes are metallic \citep{saito1998physical}, the vanishingly small bandgaps observed in the simulations are possibly a consequence of symmetry breaking arising due to numerical artifacts. In any case, given the negligible bandgaps, the results here indicate that armchair nanotubes are metallic at ambient conditions \cite{white2005fundamental}, in agreement with previous work as well as experimental measurements \cite{tans1997individual}.

\begin{figure}[htb]
\subfloat[Armchair]{\includegraphics[keepaspectratio=true,width=0.45\textwidth]
 {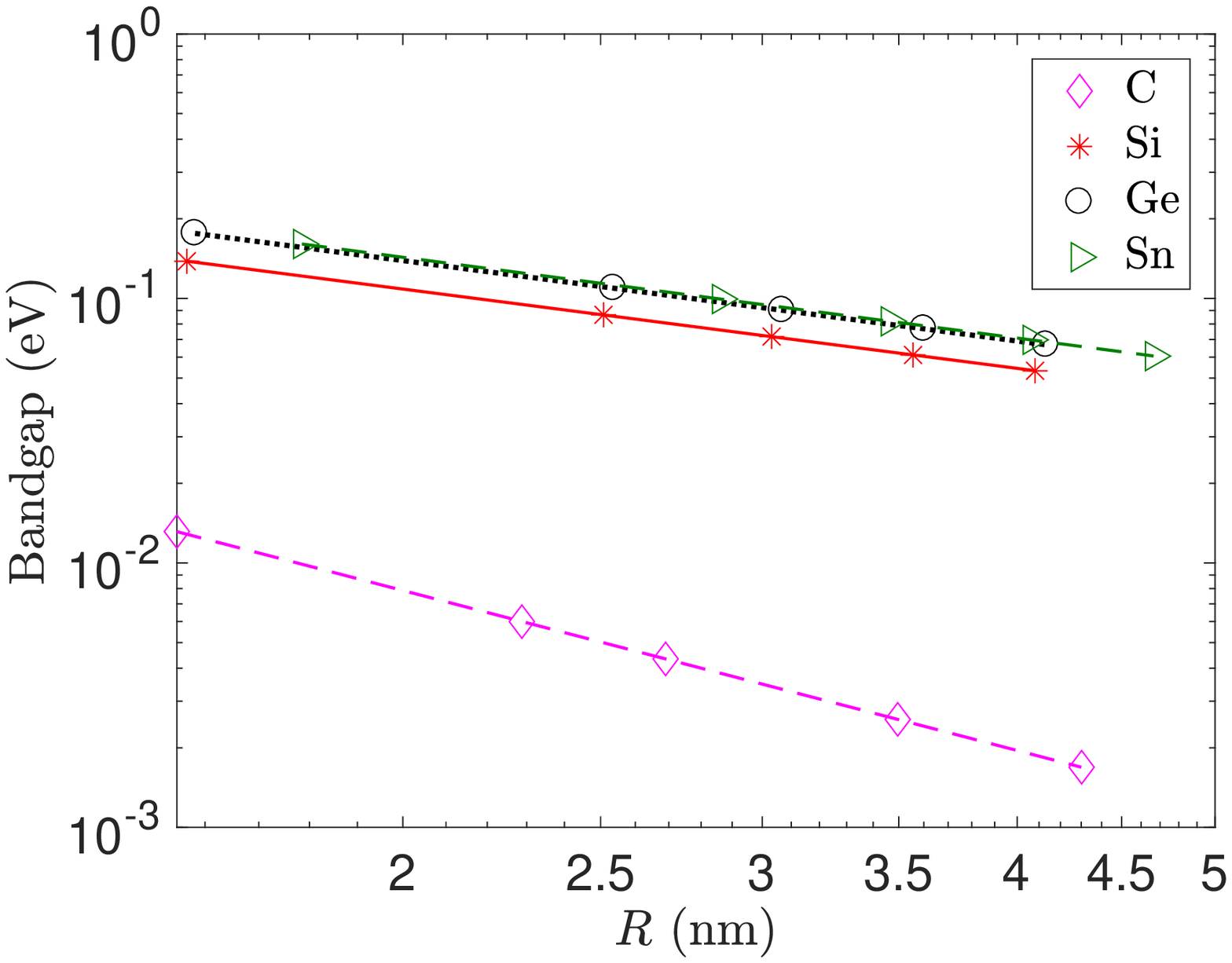}\label{Fig:Bandgap_armchair}}
\subfloat[Zigzag type I]{\includegraphics[keepaspectratio=true,width=0.45\textwidth]{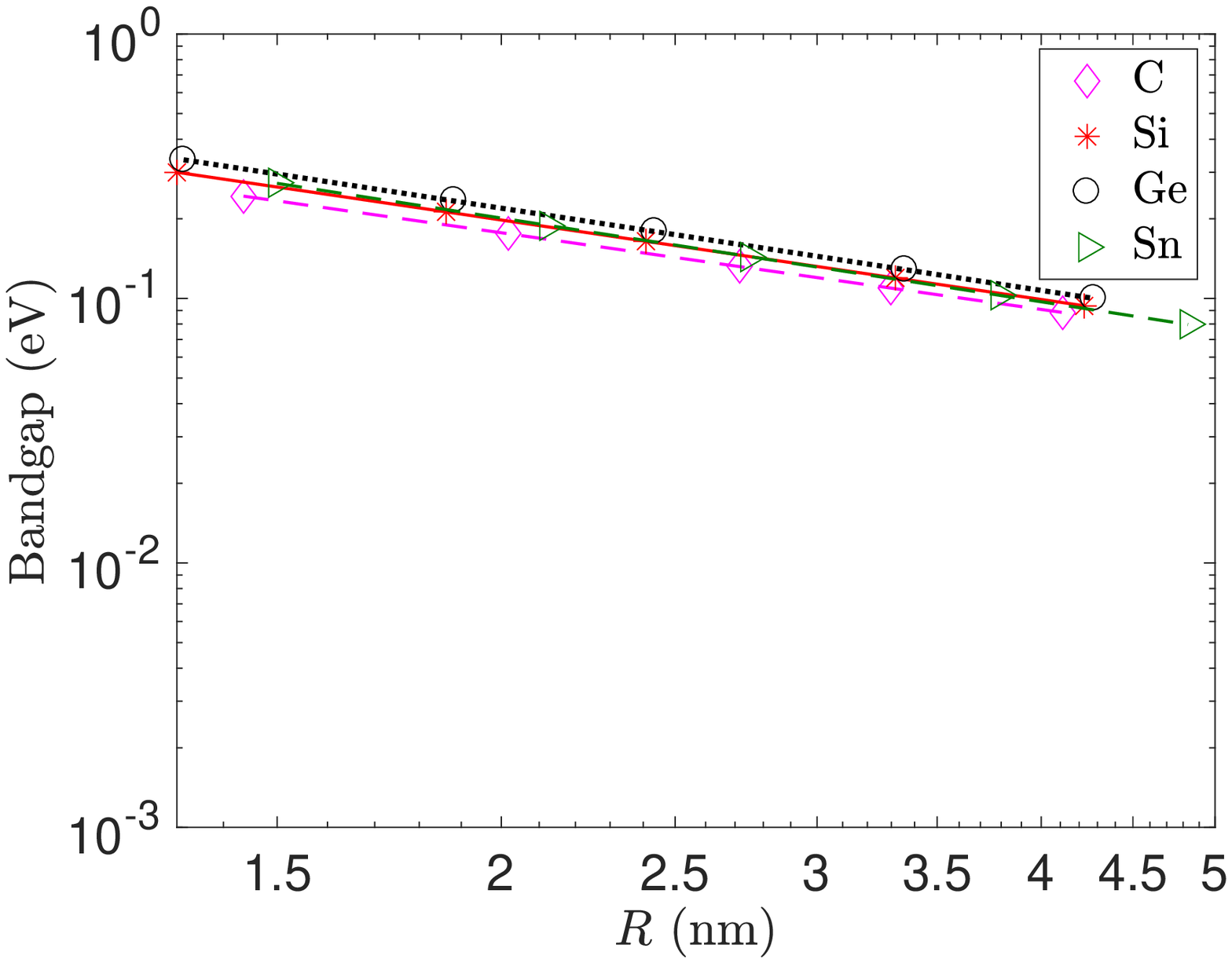}\label{Fig:Bandgap_zigzag_mod1}}\\
\subfloat[Zigzag type II]{\includegraphics[keepaspectratio=true,width=0.45\textwidth]{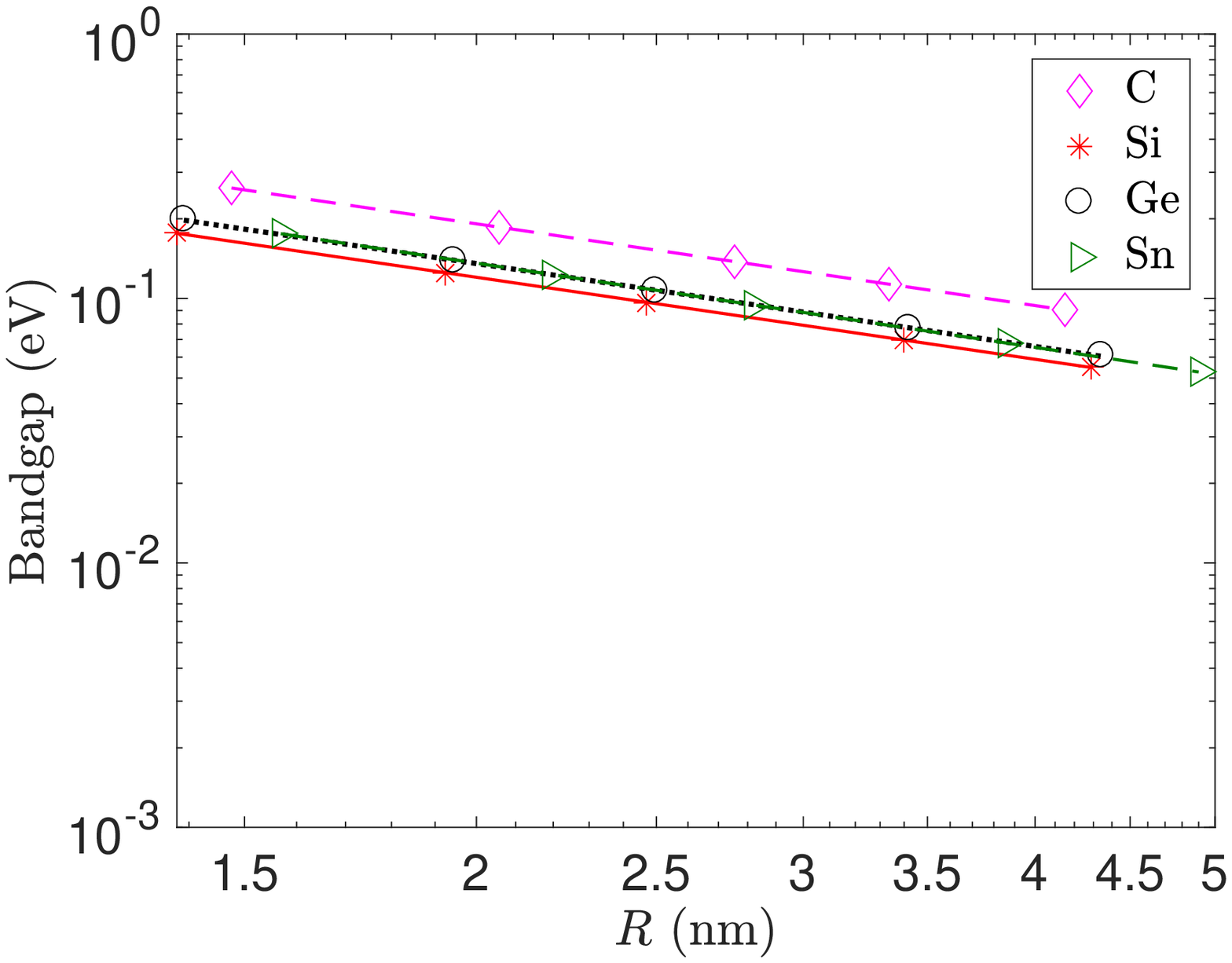}\label{Fig:Bandgap_zigzag_mod2}}
\subfloat[Zigzag type III]{\includegraphics[keepaspectratio=true,width=0.45\textwidth]{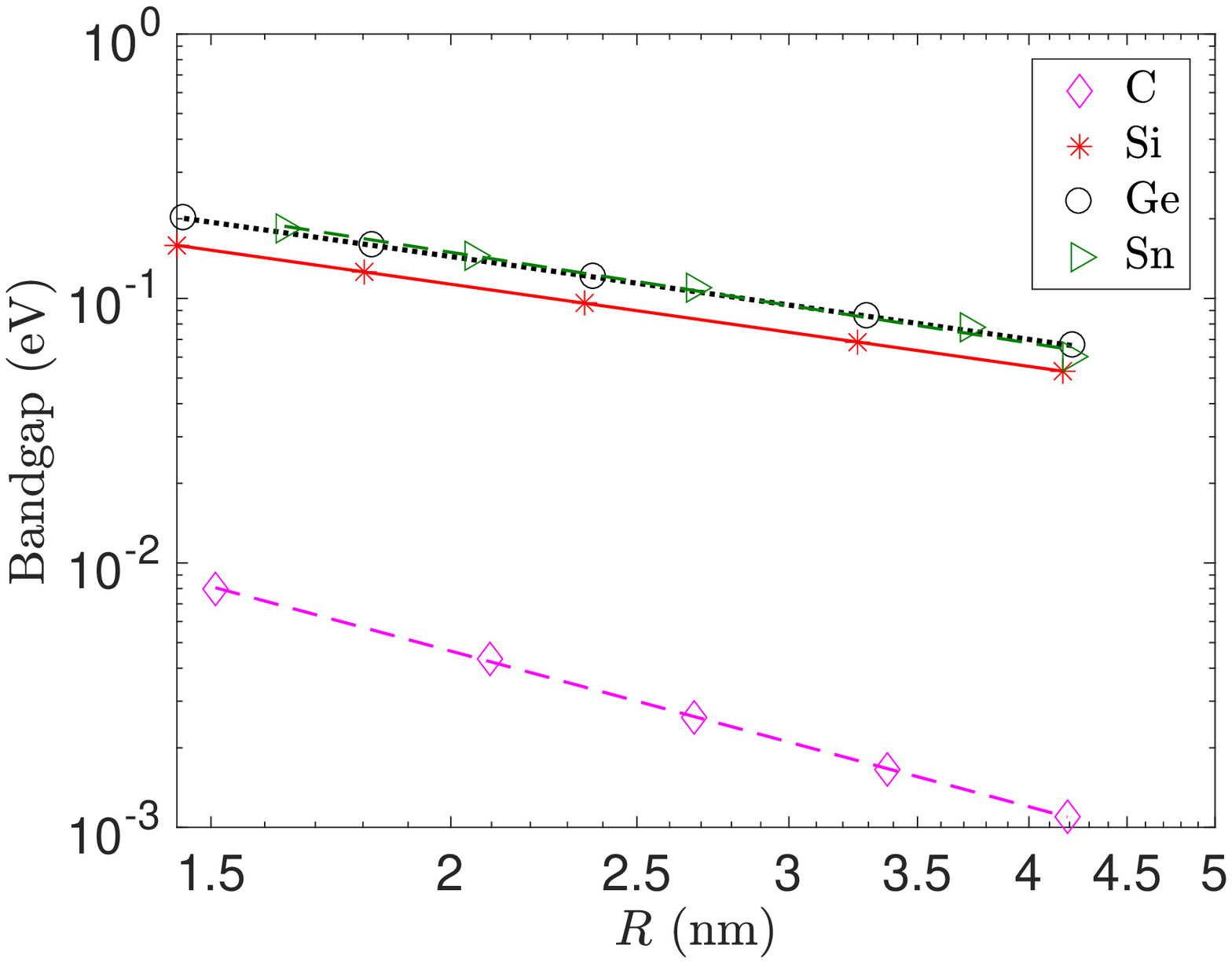}\label{Fig:Bandgap_zigzag_mod3}}
\caption{Bandgap of X (X=C, Si, Ge, Sn) nanotubes as a function of their radius $R$. The straight lines represent fits to the data.} 
\label{Fig:Bandgap_decay}
\end{figure}

Next, we determine the variation of bandgap with nanotube radius $R$, the results of which are presented in Fig.~\ref{Fig:Bandgap_decay}. Anticipating an inverse power-law dependence, we compute the decay exponents through straight line fits and present the results so obtained in Table \ref{Table:Bandgap}. It is clear that the nearly all nanotubes possess a close to inverse linear dependence with radius, the exceptions being armchair and zigzag type III nanotubes of carbon, which possess a close to inverse quadratic dependence. This atypical  dependence in carbon nanotubes is consistent with results obtained from elaborately constructed tight binding models for graphene that are able to explicitly account for curvature effects \citep{ding2002analytical}. Note that these effects are automatically incorporated into our ab-initio simulations and therefore provide an elegant route to the fitting of the material parameters that appear in such tight binding models. The minor deviation of the computed decay exponents from $1$ or $2$ is perhaps indicative that the bandgap of such materials is better expressed by a relationship of the form $ \frac{c_1}{R}+\frac{c_2}{R^2}$. However, these variations are quite small for the Xene tubes considered here, and therefore have been neglected. Overall, apart from armchair carbon nanotubes that have been previously considered as metallic, the scaling laws obtained here for the bandgap as a function of the radius are in good agreement with literature \citep{Lieber_CNT, ding2002analytical, ouyang2002fundamental,wang2017band}

\begin{table}[htb]
\centering
\begin{tabular}{ccccc}
\hline
X  &  Armchair &          & Zigzag &          \\   
      &       & Type I &  Type II &  Type III    \\
\hline
C  & -2.01 & -0.96 & -1.03  & -1.95    \\
Si  & -1.00 & -1.00 & -1.03  &  -1.03   \\
Ge & -1.01 & -1.03 & -1.04  &  -1.04   \\
Sn  & -1.02 & -1.05 & -1.06 &   -1.14 \\
\hline
\end{tabular}
\caption{Power law exponents for the decay of bandgap with radius in X  nanotubes.}
\label{Table:Bandgap}
\end{table}

It is worth noting that the LDA exchange-correlation functional has limitations in the prediction of quantitatively accurate bandgaps \citep{sham1985density, hybertsen1985first, perdew1983physical, van2006quasiparticle}. However, qualitative trends such as those discussed above are expected to be representative of the physical behavior. Indeed, the proposed formulation does not have any fundamental difficulty in dealing with  semilocal and hybrid exchange-correlation functionals, which can be employed for making more quantitatively accurate predictions, making it a worthy subject for future work.

\subsection{Bending moduli of Xene (X=C, Si, Ge, Sn) sheets} \label{subsec:sheet_bending_studies}
We now use the proposed method to calculate the bending moduli of Xene (X=C, Si, Ge, Sn) sheets along the armchair and zigzag directions. Specifically, we consider uniformly bent sheets with radii of curvature ranging from $R=1.3$ nm to $R=4.8$ nm. In order to significantly increase the efficiency of the simulations, we approximate a Xene sheet bent along the armchair or zigzag  directions as armchair or zigzag  nanotubes, respectively, with the radius of the nanotube chosen such that it matches the desired radius of curvature.\cite{Banerjee2016cyclic} This strategy indeed neglects edge related effects, which are expected to play a relatively minor role in the current context. Such an  approximation can be justified by appealing to Saint-Venant's principle \citep{iesan2006saint} and the nearsightedness of matter \citep{prodan2005nearsightedness} at the continuum and electronic structure scales, respectively. 

\begin{figure}[htb] \centering
\subfloat[Armchair]{\includegraphics[keepaspectratio=true,width=0.45\textwidth]{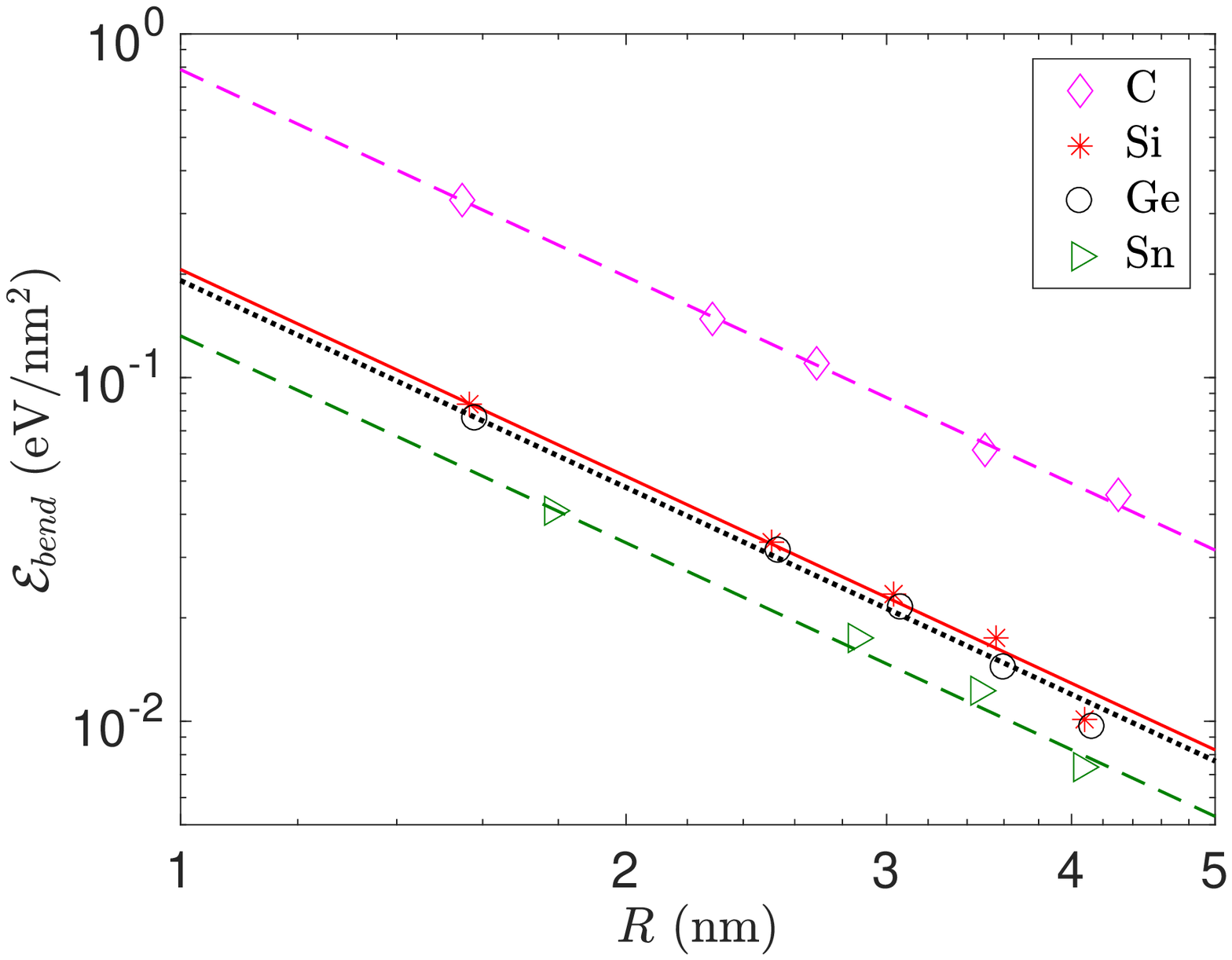}\label{Fig:BendingEnergy_armchair}}
\subfloat[Zigzag]{\includegraphics[keepaspectratio=true,width=0.45\textwidth]{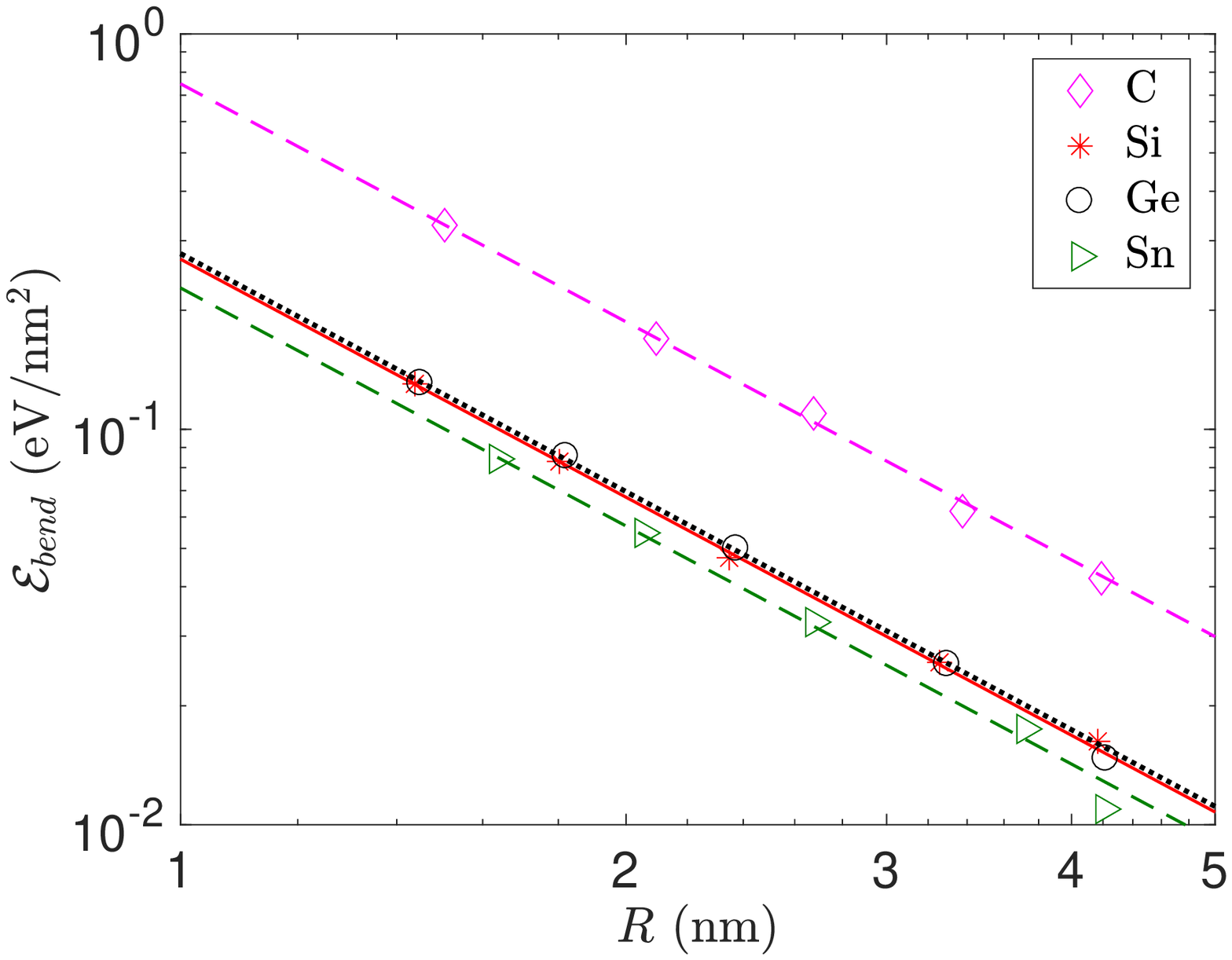}\label{Fig:BendingEnergy_zigzag}}
\caption{Bending energy $\mathcal{E}_{bend}$  of Xene (X=C, Si, Ge, Sn) sheets as a function of radius of curvature $R$. The straight lines represent fits to the data.} 
\label{Fig:Bending_Energy}
\end{figure}

Using the strategy described above, we calculate the bending energy\footnote{In the context of nanotubes, this is referred to as formation energy or strain energy.} $\mathcal{E}_{bend}$ as a function of the radius of curvature $R$ along the armchair and zigzag directions, and plot the results so obtained in Fig.~\ref{Fig:Bending_Energy}. The bending energy  $\mathcal{E}_{bend}$  at a given radius of curvature $R$ is defined to be the difference in the free energy per fundamental domain between the nanotube (with radius $R$) and sheet configurations, normalized by the area of the sheet within the fundamental domain. We observe that  $\mathcal{E}_{bend}$ has an inverse quadratic dependence on $R$, which signifies a Kirchhoff-Love type bending behavior\citep{reddy2006theory}, i.e., 
\begin{equation}
{\mathcal{E}}_{bend}(R) = \frac{1}{2} D \left(\frac{1}{R}\right)^2 \,,
\end{equation}
where $D$ can be interpreted as the modulus along the direction of bending, i.e., armchair or zigzag. This quadratic dependence on curvature is in good agreement with previous such studies for carbon nanotubes.\cite{Sanchez1999}

In Table~\ref{Table:BendingModulus}, we list the bending modulus of the Xene sheets along the armchair and zigzag directions, obtained by fitting the data in Fig.~\ref{Fig:Bending_Energy}. It is clear that the bending moduli of graphene are significantly larger than the bending moduli of the other Xenes. In particular, the bending modulus of graphene is factors of $3.3$ and $6.0$ larger than stanene in the zigzag and armchair directions, respectively. This is possibly a consequence of the short and strong bonds in graphene compared to the other Xenes, particularly stanene. We also find that, apart from graphene which is known to be close to isotropic \cite{Sanchez1999,kudin2001c,wei2012bending}, there is significant anisotropy in the bending modulus between the two directions. We correlate this anisotropy with the normalized buckled distance $\delta/a$, i.e., as $\delta/a$ increases, so does the anisotropy between the bending moduli along the two directions.

\begin{table}[htb]
\centering
\begin{tabular}{ccc}
\hline
Xene & \multicolumn{2}{c}{$D$ (eV)} \\
 & Armchair &  Zigzag  \\  
\hline
C  & 1.57  & 1.50 \\
Si  & 0.41  &  0.54 \\
Ge & 0.38  & 0.56  \\
Sn  & 0.26  & 0.46 \\
\hline
\end{tabular}
\caption{Bending modulus ($D$) for the Xene sheets in the armchair and zigzag directions.}
\label{Table:BendingModulus}
\end{table} 

It is worth noting that the bending modulus of graphene computed here ($\sim 1.5$ eV) is in good agreement with previous such DFT predictions \cite{kudin2001c,wei2012bending}. The need for ab-initio calculations is clear from the scatter ($\sim 0.8-1.4$ eV  \cite{arroyo2004finite,lu2009elastic}) in the predictions made while using empirical potentials . This need is further emphasized  by the tremendously larger ($\sim 38.5$ eV \cite{roman2014mechanical}) and therefore likely unphysical results obtained  for the bending modulus of silicene using empirical potentials. Though state of the art efficient  DFT implementations could possibly have been used to calculate the bending moduli of the Xenes studied in this work, the computational cost becomes prohibitively expensive, particularly as the radius of curvature approaches values representative of those found in experiments. Indeed, by exploiting the cyclic symmetry, the proposed formulation achieves a tremendous speedup,  enabling the extremely efficient study of such systems, as quantified in the next subsection.  


\subsection{Scaling and performance} \label{subsec:parallel_scaling}
Finally, we turn to the scaling and performance of the proposed method. We choose zigzag silicon nanotubes as representative examples for this study and use discretization parameters $h=0.5$ Bohr and $N_{\eta}=3$, which results in energy and atomic forces that are converged to within $10^{-3}$ Ha/atom and $10^{-3}$ Ha/Bohr, respectively. These accuracies are more typical of those targeted in DFT calculations, including those involving geometry optimization and molecular dynamics. Indeed, as mentioned before, significantly more stringent accuracies were targeted in the previous two subsections to ensure that the scaling relations for the X nanotubes (i.e., bandgap as a function of the nanotube radius) and bending moduli of the Xene sheets were calculated to a high degree of precision. 

\begin{figure}[htb]
\subfloat[Strong scaling]{\includegraphics[keepaspectratio=true,width=0.45\textwidth]
 {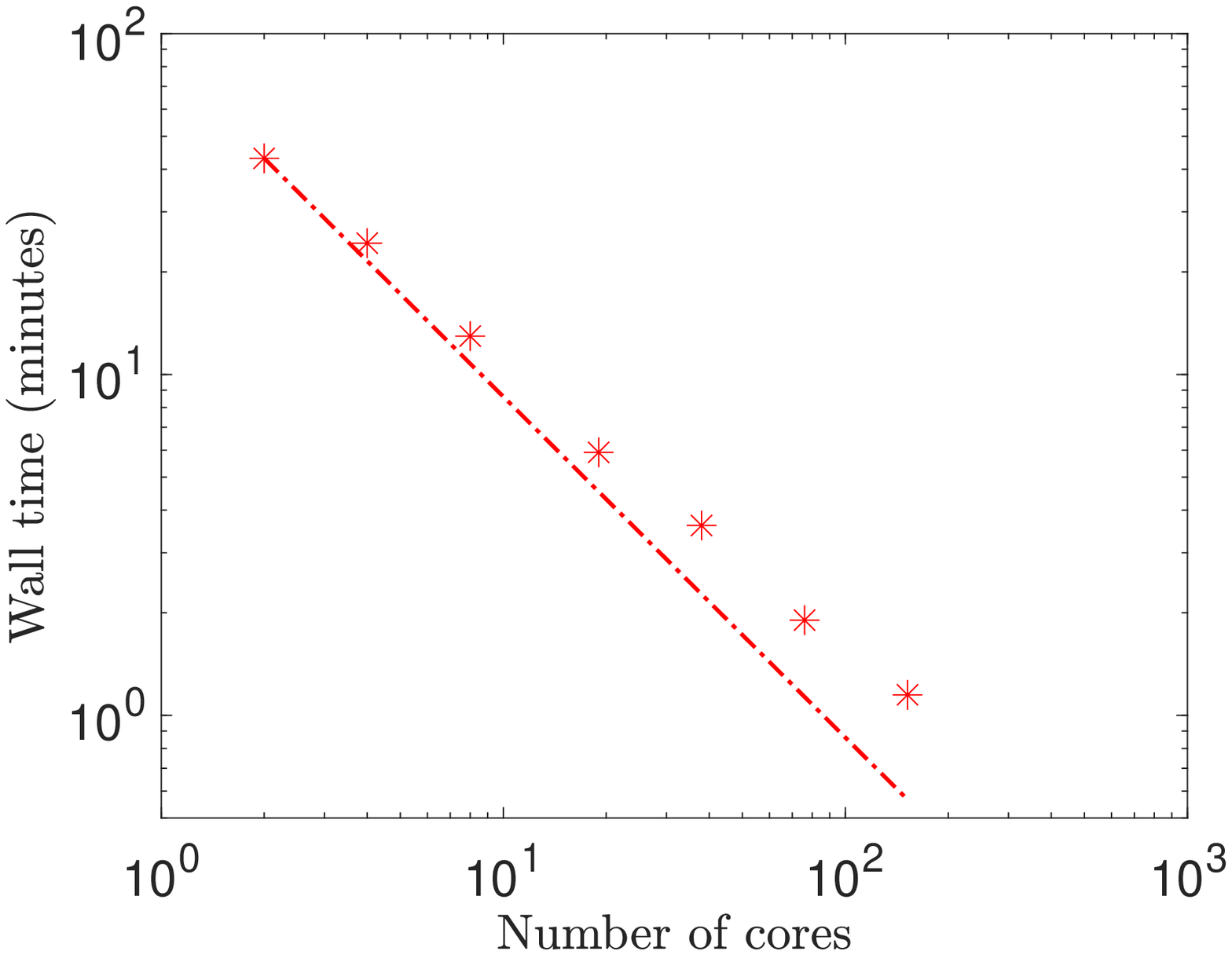}\label{Fig:StrongScaling}}
\subfloat[Weak scaling]{\includegraphics[keepaspectratio=true,width=0.45\textwidth]{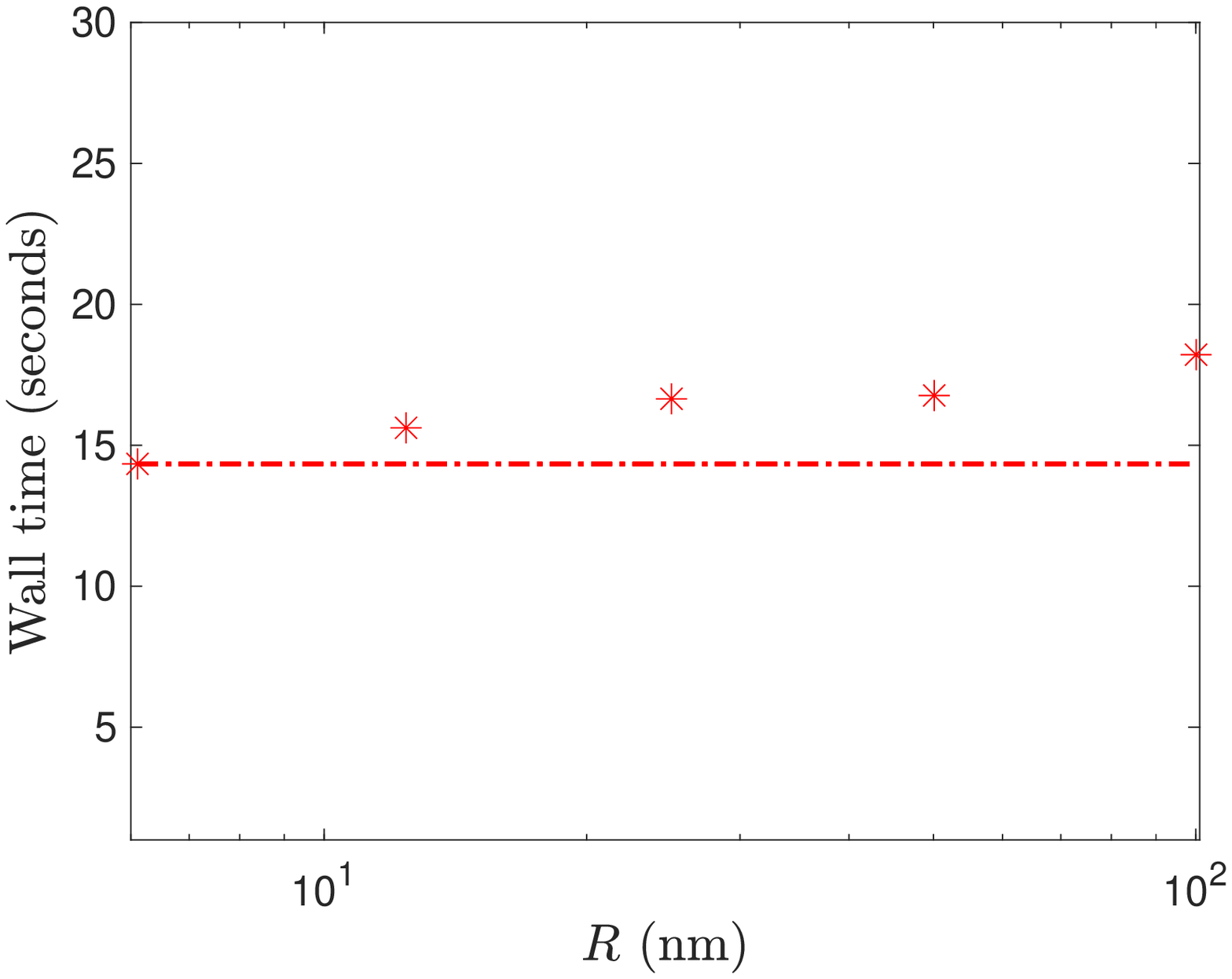}\label{Fig:WeakScaling}}
\caption{Strong and weak scaling of the proposed method for zigzag silicon nanotubes. The strong and weak scaling results correspond to wall times for the complete simulation and per SCF iteration, respectively. The strong scaling is performed for a nanotube of radius $R = 6.1$ nm. The straight lines represent ideal scaling.} 
\label{Fig:Scaling}
\end{figure}

We first perform a strong scaling study for a silicon nanotube with radius $6.1$ nm ($\mathfrak{N} = 101 $). Specifically, holding the system fixed, we increase the number of processors from $2$ to $152$ and determine the wall time associated with the complete simulation, i.e., total time for the calculation of the ground state electron density, energy, and atomic forces.  We present the results so obtained in Fig.~\ref{Fig:StrongScaling}, from which it is clear that we obtain good strong scaling, achieving an efficiency of $50\%$ on the largest number of processors relative to the smallest number of processors. The wall time on 152 processors is only 70 seconds, which is relatively small given the size of the system. Indeed, it is factor of $\mathcal{O}(100)$ smaller than SPARC, when run on the same number of cores \footnote{For this system, since SPARC faces issues with SCF convergence when the mixing parameters used in this work are chosen, the reported speedup is estimated based on the time per SCF iteration.}.  SPARC itself has been shown to be significantly more efficient compared to established planewave codes like ABINIT, highlighting the efficiency of the current approach.

Next, we perform a weak scaling study by selecting a series of silicon nanotubes with radii from $6.1$ nm ($\mathfrak{N}=101$) to $100.06$ nm ($\mathfrak{N}=1650$),  while correspondingly increasing the number of processors from $38$ to $619$. We choose $N^{proc}_D = 1$ and set $N^{proc}_K$ such that each processor works on the symmetry-adapted Hamiltonians associated with $4$ discretized characters. We plot the SCF iteration time in Fig.~\ref{Fig:WeakScaling}, from which it is clear that the proposed approach demonstrates good weak scaling for the range of systems and processors considered. Specifically, we obtain close to linear scaling, achieving $80\%$ efficiency for an increase in the nanotube radius by a factor of $\sim 16$. This can be justified by the fact that the work done per processor remains independent of system size in the proposed approach. Given that traditional DFT formulations instead scale cubically with the nanotube radius, systems such as the $100.06$ nm nanotube considered here would be exceedingly expensive, if not intractable, prior to this work.

It is worth noting that we have performed the weak scaling study by increasing the nanotube radius (i.e., $\mathfrak{N}$), while holding the system size within the fundamental domain fixed. Alternatively, we could have increased the system size within the fundamental domain, while holding the value of $\mathfrak{N}$ fixed.  However, the results obtained in this case would be similar to those obtained by the underlying SPARC code \cite{Ghosh2017cluster,Ghosh2017extended}, and therefore are not reproduced here for the sake of brevity. It is also worth noting that, unlike in the strong scaling study where we report the wall time for the complete simulation, we report the wall time per SCF iteration for the weak scaling study. This is because of the increase in number of SCF iterations with the radius of the nanotube, a behavior that can be attributed to the change in electronic properties with system size.

\begin{figure}[htb]\centering
\includegraphics[keepaspectratio=true,width=0.45\textwidth]{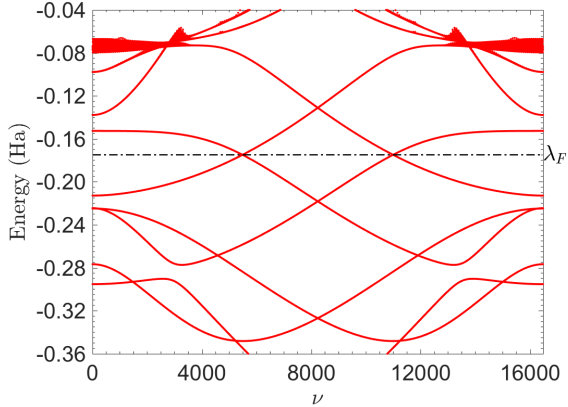}
\caption{\label{Fig:BandLargeNu} Band structure diagram for zigzag type III silicon nanotube of radius $R \sim 1$ $\mu$m along $\frac{\eta H}{2 \pi} = 0$ in $(\nu,\eta)$ space.} 
\end{figure}

The above results suggest that with sufficient computational resources, the proposed formulation allows for accurate DFT simulations of extremely large radius nanotubes, with modest wall times. To demonstrate this capability, we simulate a zigzag type III silicon tube of radius $R \sim 1$ $\mu$m ($\mathfrak{N} = 16,473$) in $53$ minutes of wall time on just $353$ processors. We have found that the system has a negligible direct bandgap of $\mathcal{O}(10^{-4})$ eV at $\frac{\eta H}{2 \pi} = 0$, consistent with the results and scaling law obtained for silicon nanotubes in Section~\ref{subsec:nanotube_electronic_studies}, further verifying the accuracy of the simulation. Fig.~\ref{Fig:BandLargeNu} shows the band structure diagram for $\frac{\eta H}{2 \pi} = 0$, which bears a high degree of resemblance to the corresponding band structure diagram of the silicene sheet, as is to be expected, given the extremely large radius of the tube. To the best of our knowledge, this is the first example in literature of a fully resolved Kohn Sham DFT calculation of a system in which one of the length scales is of the order of a micron. Indeed, this example is slightly contrived since similar results could have been obtained by traditional DFT implementations at substantially reduced cost by using a flat sheet approximation of this nanotube, given its tremendously large radius and insignificant curvature induced effects. However, it serves to demonstrate the capabilities of the proposed method, with potential application to naturally large radius nanotubes \cite{shin2004formation,mcgary2006magnetic,macak2008mechanistic}, where curvature induced effects are likely to be substantial.

In this work, though we have studied nanotube systems containing only 4 atoms in the fundamental domain, the proposed method is not restricted by this number, and given sufficient computational resources, the developed implementation can study systems containing up to a thousand atoms in the fundamental domain. Indeed, due to the significantly increased cost in such cases, the efficiency of the code would benefit from band parallelization as well as the AAR \cite{pratapa2016anderson,suryanarayana2019alternating} and DDBP \cite{xu2018discrete} methods. Even larger systems containing tens of thousands of atoms in the fundamental domain will then become accessible using the Complementary Subspace method \cite{banerjee2018two}. Such advances will enable a number of applications, including the study of nanofilm bending \cite{park2014highly,haque2003strain,huang2005bending}, which are intractable using traditional DFT methods.

\section{Concluding remarks} \label{Sec:Conclusions}
In this work, we have developed a symmetry-adapted real-space formulation of Kohn-Sham DFT for cylindrical geometries and applied it to the study of large X (X=C, Si, Ge, Sn) nanotubes. Specifically, we have started from the original Kohn-Sham equations that are posed on all of space, and reduced them to the fundamental domain by accounting for the cyclic and periodic symmetries present in the angular and axial directions of the cylinder, respectively. We have implemented this approach for parallel computations using the high-order real-space finite-difference method, and verified its accuracy with respect to established planewave and real-space codes. We have used this implementation to study the band structure properties of X nanotubes and bending properties of Xene sheets. Specifically, we have first shown that zigzag and armchair X nanotubes with radii in the range of $1$ to $5$ nm are semiconducting, other than the armchair and zigzag type III carbon variants, for which we find a vanishingly small bandgap, indicative of metallic behavior. In particular, we have found that apart from armchair and zigzag type III carbon nanotubes, which demonstrate an inverse quadratic dependence of the bandgap with respect to radius, all other nanotubes demonstrate an inverse linear dependence. Next, we have exploited the the connection between cyclic symmetry and uniform bending deformations to calculate the bending moduli of Xene sheets in both zigzag and armchair directions for radii of curvature up to $5$ nm. We have found that the sheets obey Kirchhoff-Love type bending, with graphene and stanene demonstrating the largest and smallest moduli, respectively. In addition, apart from graphene, the sheets demonstrate significant bending  anisotropy, with larger moduli along the armchair direction. Finally, we have shown that the proposed method is highly efficient and extremely well suited for parallel computations, which enables ab initio simulations of  unprecedented size for systems with a relatively large degree of cyclic symmetry. In particular, we have shown that nanotubes with radii even at the micrometer scale can be simulated with modest computational resources and effort. 

Overall, the proposed method provides an efficient framework for ab-initio simulations of 1D nanostructures with large radii as well as 1D/2D nanostructures under uniform bending, which are intractable using traditional formulations and implementations of DFT.  This opens an avenue for the ab-initio study of the flexoelectric effect \cite{hong2013first}, in which a number of open questions remain \cite{wang2019flexoelectricity}. The extension of the proposed method to include helical symmetry will enable the efficient ab-initio study of chiral nanotubes as well other nanostructures with helical symmetry, making it a worthy subject of future research.

\begin{acknowledgments}
S.G. acknowledges support from the Army Research Laboratory which was accomplished under Cooperative Agreement Number W911NF-12-2-0022. A.S.B acknowledges support from the Scientific Discovery through Advanced Computing (SciDAC) program funded by U.S. Department of Energy, Office of Science, Advanced Scientific Computing Research and Basic Energy Sciences, while at the Lawrence Berkeley National Laboratory. A.S.B also acknowledges support from the Minnesota Supercomputing Institute (MSI) for some of the computational resources that were used in this work. P.S. gratefully   acknowledges   the   support   of the National Science Foundation (CAREER-1553212). This research was supported in part through research cyberinfrastructure resources and services provided by the Partnership for an Advanced Computing Environment (PACE) at the Georgia Institute of Technology, Atlanta, Georgia, USA. Some of the computations presented here were conducted on the Caltech High Performance Cluster partially supported by a grant from the Gordon and Betty Moore Foundation. The authors acknowledge the valuable comments and suggestions of the anonymous referees.

\end{acknowledgments}


\end{document}